\documentclass[twocolumn,dvipsnames]{aastex63}
\usepackage{graphicx} % Required for inserting images
\usepackage{multirow}
\usepackage{amsmath}
\usepackage{bm}
\usepackage{lineno}

\newcommand{\eq}[2]{\begin{equation} \label{eq:#1} #2 \end{equation}}
\newcommand{\Eref}[1]{(\ref{eq:#1})}

\newcommand{\edits}[1]{#1}

\begin{document}
%\linenumbers

\title{QPOs in compact sources as a non-linear hydrodynamical resonance: Determining spin of compact objects}

%%%%%%%%%% Authors of the paper %%%%%%%%%%%%%%%%
\author[0000-0001-8451-0806]{Arghya Ranjan Das$^{\dagger}$}
\email{arghyadas@iisc.ac.in$^{\dagger}$}
\affiliation{Department of Physics, Indian Institute of Science, Bangalore 560012, India}

\author[0000-0002-3020-9513]{Banibrata Mukhopadhyay$^{\ddagger}$}
\email{bm@iisc.ac.in$^{\ddagger}$}
\affiliation{Department of Physics, Indian Institute of Science, Bangalore 560012, India}

%%%%%%%%%% Abstract of the paper %%%%%%%%%%%%%%%%
\begin{abstract}
Origin of wide varieties of quasi-periodic oscillation (QPO) observed in compact sources is still not well established. Its frequencies range from mHz to kHz spanning all compact objects. Are different QPOs, with different frequencies, originating from different Physics?  
We propose that the emergence of QPOs is the result of  
nonlinear resonance of fundamental modes present in accretion disks forced by external modes including that of the spin of the underlying compact object. Depending on the properties of accreting flow, e.g. its velocity and gradient, resonances, and any mode locking, take place at different frequencies, exhibiting low to high frequency QPOs. We explicitly demonstrate the origin of higher frequency QPOs for black holes and neutron stars by a unified model and outline how the same physics could be responsible to produce lower frequency QPOs. The model also predicts the spin of black holes, and constrains the radius of neutron stars and the mass of both.    
\end{abstract}

%\keywords{black hole physics; stars: neutron; gravitation;  hydrodynamics; accretion, accretion disks; stars: oscillations}

%%%%%%%%%%%%%%%%%%%%%%%%%%%%%%%%%%%%%%%%%%%%%%%%%%

%%%%%%%%%%%%%%%%% BODY OF PAPER %%%%%%%%%%%%%%%%%%

\section{Introduction}
Since its first discovery in the 1980s \citep{motch1983simultaneous, vanderKlis1985}, there are numerous discoveries of quasi-period oscillations (QPOs) with their varieties of properties in black hole and neutron star sources, even sometimes in white dwarfs (e.g., \citealt{Mauche_2002, Titarchuk_2002,2002MNRAS.333..411W}). While for neutron stars, QPOs are mostly observed in the order of kilo Hz (kHz) frequencies, for black holes they vary from the fraction of Hz to a faction of kHz, the latter is called the high frequency (HF) QPO. While HF QPOs are observed mainly in the high-soft (HS) state of a black hole system \citep{Belloni_2012}, the low frequency QPOs are mostly observed in harder states \citep{Belloni_2012, 10.1111/j.1365-2966.2012.22037.x, 10.1093/mnras/stw3363}. Moreover, often the source exhibits jets along with low frequency QPOs being in a hard state, e.g. in a canonical low-hard (LH) state. 

Often the kHz/HF QPOs appear with a pair. It had been argued and apparently observed that the separation between the kHz QPO frequencies in a pair is similar in order of magnitude to the half of the spin frequency or the spin frequency itself of the fast or slowly rotating neutron stars respectively \citep{van_der_Klis_2000, Kluźniak_2004}. This argues for a relation between kHz QPOs and the spin of neutron stars. On the other hand, the HF QPOs in a pair for black holes were argued to be in a $3:2$ ratio \citep{10.1046/j.1365-8711.2003.07018.x,doi:10.1146/annurev.astro.44.051905.092532,10.1093/mnrasl/slv196}, but also see \citealt{Homan_2003}, whose observational analysis argued against it, even though some authors argued QPOs from the neutron star, Sco X-1, appear in this ratio \citep{Abramowicz_2003,REBUSCO2008855}. Alternative viewpoints had been put forth suggesting  against the fixed $3:2$ ratio, and instead suggesting a multi-peaked distribution of QPO frequencies for Sco X-1 and other sources, namely, 4U 1608–52, 4U 1636-53, 4U 1728-34, and 4U 1820-30, with the ratio of high to low frequency QPOs to be 3/2, 4/3, 5/4, 7/5 and 9/7 spanning around sources, not all ratios appearing in all sources though (\citealt{belloni2005distribution, 10.1111/j.1365-2966.2007.11486.x}). 

Now the question arises, could all the QPOs, particularly the kHz/HF ones, be originated by the same mechanism? Generally, the QPOs are expected to be processed in the matter inflowing towards compact objects, i.e. in the accretion disk therein, probably in its inner region where the effect of gravity is stronger. Therefore, it is expected to be originated from the same/similar mechanism, broadly independent of the nature of the compact object, which is primarily controlled by gravity. Therefore, we target explaining the origin of (kHz/HF) QPOs by a unified mechanism in an accretion flow. 

Often kHz/HF QPOs are argued to be involved with parametric resonance phenomena and related mode locking \citep{Stella:1997tc,Stella:1999sj,Lamb_2001, Cadez:2008iv, Kostic:2009hp, Germana:2009ce, Kluzniak:2002bb, Abramowicz:2003xy, Rebusco:2004ba, Nowak:1996hg, Torok:2010rk, Torok:2011qy, Kotrlova:2020pqy}.
%\ref{abramowicz, miller, lamb etc.} 
\cite{Mukhopadhyay_2009} initiated modeling, both, kHz QPOs of neutron stars and HF QPOs of black holes in a unified scheme. The author argued, based on a very schematic model, the kHz/HF QPOs to be the result of higher order nonlinear processes in accretion disks. Based on this QPO model, the spin of black holes and neutron stars (if unknown) was also estimated. Since QPO carries the information of spacetime around the compact object, it can also be used to test any modification to gravity such as asymptotically flat modified $f(R)$-gravity \citep{Kalita_Bani, Das2022} in strong gravity regime.

Arguably the HF QPOs are the most reliable source of spin measurement once the correct model is known \citep{doi:10.1146/annurev.astro.44.051905.092532}. 
There are many avenues in which black hole's spin measurement can be done such as:
\edits{
polarimetry \citep{1980ApJ...235..224C, 1975ApJ...198L..73L}, continuum fitting \citep{Zhang_1997, Dovčiak_2004, Davis_2005, Li_2005, 10.1111/j.1365-2966.2011.18446.x}, the Fe K line \citep{REYNOLDS2003389, Reynolds_2008_1}, and HFQPOs (e.g. \citealt{Mukhopadhyay_2009}).
}
Arguably, the best current method, continuum fitting, has a drawback that it requires precise estimations of parameters such as black hole mass $(M)$, disk inclination $(i)$, and distance. On the other hand, assuming that we have a well tested QPO model, spin estimation would require only only the estimate of $M$. In fact our current model constraints the already existing mass range further to a more precise estimation from the observed QPO frequencies. 

Earlier it has been shown that QPO may arise due to non-linear resonance phenomenon in accretion disk placed in strong gravity \citep{Kluźniak_2004, refId0, Blaes_2007}. The hydrodynamic/magnetohydrodynamic equations that govern the accretion dynamics and disk structure are non-linear, hence a non-linear response is expected. In the present work, we propose a modification to the epicyclic frequencies due to motion of fluid (instead of test particle) around the compact object. We consider these modified frequencies as fundamental modes to model QPO frequencies based on non-linear dynamics and resonance following the approach of \cite{Mukhopadhyay_2009}. The model successfully describes the observational trends in QPOs and is able to reproduce the observed frequencies. In the process of reproducing observed results, the model estimates the mass $(M)$ and spin parameter $(a)$ of the black holes and radius of neutron stars, and also the estimate of spin frequency of neutron star (if unknown) is made. 

The plan of the paper is as follows. In the next section we derive the modified epicyclic frequencies for fluids around a compact object. Section 3 devotes outlining the basic properties of nonlinear resonance and how does it govern QPOs in the accretion disk. Subsequently, we outline the basic characteristics of black holes and neutron stars and their basic parameters required for our model in section 4. The next two sections, 5 and 6,
explore our model to explain observed QPOs for neutron stars and black holes respectively. Before ending, we outline in section 7 showing briefly that our model could explain other QPO frequencies as well. Finally, we end with a summary and conclusions in section 8.

\section{Modified epicyclic frequency for fluid} \label{sec:mod_ep}
%In literature the fluid dynamics can be explained using Navier-Stokes equation. But for in-viscid fluid Euler equation suffice for any analysis. The Euler equation in fluid mechanics takes the form,
We consider a small annulus region of an accretion disk. This small region is assumed to be incompressible and the momentum balance is given by
\eq{NV_eq}{\frac{\partial \textbf{u}}{\partial t}+(\textbf{u}\cdot\nabla)\textbf{u}+\nabla\left(\frac{P}{\rho_0}\right) = \textbf{F},}
where $\textbf{u}$, $P$ and $\rho_0$ are the velocity, pressure and density of the fluid respectively and $\textbf{F}$ is the external force per unit mass including the contribution due to gravity.  

Now consider a fluid packet to be moving at a radius $\textbf{r}$. Due to several possible sources of disturbances in general, we consider a small orbital perturbation of the fluid packet as
\eq{perturb}{\textbf{r}\rightarrow\textbf{r}+\delta\textbf{r}.}
Assuming locally polytropic equation of state, from equations \Eref{NV_eq} and \Eref{perturb}, the linearly perturbed equation takes the form of
\eq{perturb_NV}{\frac{\partial \delta\textbf{u}}{\partial t}+(\delta \textbf{u}\cdot\nabla)\textbf{u}+(\textbf{u}\cdot\nabla)\delta\textbf{u} = \delta \textbf{F},}
where the quantities under $\delta$ imply the perturbation of the original variables.
Combining the first and the third terms from equation \Eref{perturb_NV}, we have
\eq{perturb_NV_2}{\frac{d \delta\textbf{u}}{d t}+(\delta \textbf{u}\cdot\nabla)\textbf{u} = \delta \textbf{F}.}
This equation effectively describes a forced damped oscillator. 
We further confine the equatorial accretion disk around compact objects assuming axisymmetry. Hence, the second term in equation \Eref{perturb_NV_2} in cylindrical coordinate system can be broken down to its components as
\eq{2}{(\delta \textbf{u}\cdot\nabla)\textbf{u} = \left[\delta u_r\left(\frac{\partial u_r}{\partial r}\right)-\left(\frac{u_\phi}{r}\right)\delta u_\phi\right]\hat{r}+\delta u_r\left(\frac{\partial u_\phi}{\partial r}\right)\hat{\phi}.}
Now we should note that $u_\phi = r\dot{\phi}\equiv r\Omega_\phi$ and thus due to separation of orbits we can write
\begin{eqnarray}
\nonumber
    \frac{d\delta u_\phi}{dt} &=& \frac{d}{dt}(r\delta\dot{\phi}+\Omega_\phi \delta r)\\ 
    \label{eq:4}
    &=&r\delta\ddot{\phi}+\Omega_\phi\delta\dot{r},
\end{eqnarray}
assuming effectively circular orbit when $u_r=\dot{\Omega}_\phi\sim 0$. Thus from equations \Eref{perturb_NV_2}, \Eref{2} and \Eref{4}, the components of the perturbed momentum balanced equation can be written as
\begin{eqnarray}
        \label{eq:R_series}
        \delta\ddot{r}+\gamma\delta\dot{r}-\Omega_\phi(r\delta\dot{\phi}-\Omega_\phi \delta r) &= (\delta\textbf{F})_r,\\
        \label{eq:z_series}
        \delta\ddot{z} &= (\delta\textbf{F})_z,\\
        \label{eq:phi_series}
        r\delta\ddot{\phi}+\Omega_\phi\delta\dot{r}+\xi\delta\dot{r} &= (\delta\textbf{F})_\phi,
\end{eqnarray}
where $\xi = r\Omega_{\phi,r}$ and $\gamma = {\partial u_r}/{\partial r}$.

Now we consider the change (or the perturbation) in the force due to two reasons, one is due to the orbital perturbation and the other due to the presence of some forcing 
in the system. Thus the combined perturbed force will take the form as
\eq{}{\delta\textbf{F} = \textbf{F}_{forced}+\delta\textbf{F}_{orbit},}
%The orbital part of the perturbed force can be written in the components in the cylindrical coordinates as
%\eq{}{\delta\textbf{F}_{orbit}= F_r\hat{r}+F_z\hat{z}+F_\phi\hat{\phi}.}
where the change in the orbital force in the axisymmetric equatorial system becomes 
\eq{orbit_force_term}{
\delta\textbf{F}_{orbit}= -\left( \Omega_r^2\delta r\right)\hat{r}-\left(\Omega_z^2\delta z\right)\hat{z}.}
The term $\Omega_j$ can be written as $\Omega_j = 2\pi\nu_j$ such that the frequencies $\nu_r$ and $\nu_z$ are the epicyclic frequencies and $\nu_\phi$ is the orbital frequency of a test particle moving in the equatorial circular orbit around the compact object. In the Kerr metric, these frequencies  are given as 
\begin{eqnarray}
    \nu_\phi &=& \frac{1}{2\pi}\frac{1}{r^{3/2}+a}\frac{c^3}{GM},\label{nuphi}\\
    \nu_r &=& \frac{\nu_\phi}{r}\sqrt{\Delta-4(\sqrt{r}-a)^2},\label{nur}\\
    \nu_z &=& \frac{\nu_\phi}{r}\sqrt{r^2-4a\sqrt{r}+3a^2},\label{nuz}
\end{eqnarray}
where $\Delta=r^2-2r+a^2$, $M$ is the mass of central compact object, $G$ the Newton's gravitational constant and $c$ the speed of light.
\noindent For simplicity, we take the force to be varying with some frequency $\omega$, of the form,
\eq{forced_term}{\textbf{F}_{forced} = (F_{0r}\hat{r}+F_{0z}\hat{z}+F_{0\phi}\hat{\phi})\cos{\omega t}.}

\noindent Hence, the following equations of perturbation can be written from equations \Eref{R_series}, \Eref{z_series}, \Eref{phi_series}, \Eref{orbit_force_term} and \Eref{forced_term} as
\begin{eqnarray}
    && \delta\ddot{r}+\gamma\delta\dot{r}-\Omega_\phi(r\delta\dot{\phi}-\Omega_\phi \delta r) = -\Omega_r^2\delta r+F_{0r}\cos{\omega t},\nonumber \\ \label{eq:r_eq1}\\ 
    \label{eq:z_eq1}
    &&\delta\ddot{z} = -\Omega_z^2\delta z+F_{0z}\cos{\omega t},\\
    \label{eq:phi_eq1}
    &&r\delta\ddot{\phi}+\Omega_\phi\delta\dot{r}+\xi\delta\dot{r} = F_{0\phi}\cos{\omega t}.   
\end{eqnarray}
The above equations are similar to the components of  equation describing the forced damped harmonic oscillator. Now as a solution technique, we choose the force $F_{0j}\cos{\omega t}$ as $F_{0j}e^{i\omega t}$ (actually real part of it), where $j\in\{r,z,\phi\}$, and we make the ansatz that all the perturbations will have the form
\begin{eqnarray}
   \label{eq:r_ansatz}
    \delta r =& \mathcal{R}(\omega)e^{i\omega t},\\
    \label{eq:z_ansatz}
    \delta z =& \mathcal{Z}(\omega)e^{i\omega t},\\
    \label{eq:phi_ansatz}
    \delta \phi =& \Phi(\omega)e^{i\omega t}. 
\end{eqnarray}
Putting equations \Eref{r_ansatz}, \Eref{z_ansatz} and \Eref{phi_ansatz} in equations \Eref{r_eq1}, \Eref{z_eq1} and \Eref{phi_eq1}, we obtain three equations for $\mathcal{R}(\omega)$, $\mathcal{Z}(\omega)$ and $\Phi(\omega)$ which can be solved. As a first case, for simplicity, we assume $F_{0r}=F_{0z}=F_{0\phi}=F_0$. Having this, the solution for the amplitudes can be found to be of the form
\begin{eqnarray}
    \mathcal{R}(\omega)&=& \frac{F_0 \left(\omega -i \Omega _{\phi
   }\right)}{\omega  \left[\Omega _{\phi } \left(\xi +2
   \Omega _{\phi }\right)+i \omega  (\gamma +i \omega )+\Omega
   _r^2\right]}\\
   \mathcal{Z}(\omega)&=&-\frac{F_0}{\omega ^2-\Omega
   _z^2}\\
   \Phi(\omega)&=&-\frac{F_0 \left[-i \omega  (\xi -\gamma -i
   \omega )+\Omega _r^2-i \omega  \Omega _{\phi }+\Omega
   _{\phi }^2\right]}{\omega^2 r\left[\Omega _{\phi }
   \left(\xi +2 \Omega _{\phi }\right)+i \omega  (\gamma +i
   \omega )+\Omega _r^2\right]}.\nonumber\\
\end{eqnarray}
The absolute values of the amplitudes (i.e. those up to a phase factor), are given by

\begin{eqnarray}
    |\mathcal{R}(\omega)| &=& \frac{|F_0|\sqrt{\omega^2+\Omega_\phi^2}}{\omega\sqrt{\gamma^2 \omega^2+\left[\Omega_{\phi } \left(\xi +2 \Omega_{\phi}\right)+\Omega_r^2-\omega^2\right]^2}},\\
    |\mathcal{Z}(\omega)|&=&\frac{|F_0|}{\omega ^2-\Omega
   _z^2},\\
    |\Phi(\omega)|&=&\frac{\left| F_0\right|  \sqrt{\omega ^2 \left(\xi -\gamma +\Omega_{\phi }\right)^2+\left(\Omega_r^2-\omega ^2+\Omega _{\phi
   }^2\right)^2}}{\omega^2  r  \sqrt{\gamma^2 \omega^2+\left[\Omega _{\phi } \left(\xi +2 \Omega _{\phi }\right)+\Omega
   _r^2-\omega ^2\right]^2}.}\nonumber\\
\end{eqnarray}
Therefore, the perturbation $\delta z$ is maximized when we have
\eq{mod_epi1}{\omega = \Omega_z,}
and $\delta r$ and $\delta\phi$ are maximum when 
\eq{maxcond}{f(\omega) =\gamma^2 \omega ^2+\left[\Omega_{\phi } \left(\xi +2 \Omega _{\phi }\right)+\Omega_r^2-\omega ^2\right]^2}
is minimum. It can be easily found that $f(\omega)$ is minimized for 
$\gamma=0$ (for a circular orbit) when \eq{mod_epi2}{
\omega = \sqrt{r\Omega_{\phi,r}  \Omega _{\phi }+\Omega _r^2+2 \Omega _{\phi }^2}.}
Thus from equations \Eref{mod_epi1} and \Eref{mod_epi2} we obtain the modified epicyclic frequencies for fluid moving in an accretion disk in the equatorial circular orbit, hence in a Keplerian accretion disk, denoted as $\nu_1$ and $\nu_2$, given by
\begin{eqnarray}
    \nu_1 &=&  \sqrt{\nu _r^2+2\nu _{\phi }^2+r\nu_{\phi,r}  \nu _{\phi }},\label{eq:mod_QPO_main_1}\\
    \nu_2 &=& \nu_z. \label{eq:mod_QPO_main_2}
\end{eqnarray}
This modification can be summarized to be that the vertical epicyclic frequency remains unchanged as given by that of a test particle, as in $\nu_2$. However, there is a combination of the radial epicyclic frequency and the orbital frequency, as defined for a test particle, leading to a new epicyclic frequency for fluid as $\nu_1$. 
\section{Nonlinear resonance and QPO}
In the nonlinear regime, a perturbation at the stellar spin frequency $\nu_s$ can weakly excite a disk motion at the same frequency. This is expected as a rotating central object may influence the disk fluid. If a mode frequency is already present along with the perturbing frequency, another mode can be excited at a combination of the frequencies  \citep{LL1976, Abramowicz:2003xy}. In addition, if the frequency difference of two said modes excited in a system is close to $p\nu_s$, a resonance may occur such that the system is responsive
to this frequency difference only (where $p$ is a number such as $n/2$ and $n$ is an integer, for instance 1 or 2 in the present model). This resonance leads to mode-locking. Indeed some neutron star systems show that the frequency difference between twin kHz QPOs remains close to the stellar spin frequency or the half the spin frequency, even though the QPO frequencies vary over time \citep{van_der_Klis_2000}.

The underlying concept is that when a compact object rotates, it creates a new mode in the surrounding region corresponding to the spin frequency $\nu_s$ which couples to those corresponding to $\nu_1$ and $\nu_2$, already existing in the disk. This coupling will lead to the emergence of new modes with frequencies $\nu_{1,2}\pm p\nu_s$. Now at a certain radius if the difference of $\nu_1$ and $\nu_2$ is close to zero and, additionally, if $\nu_s/2$ (for nonlinear regime) or $\nu_s$ (for linear regime) is close to the frequency difference of any two newly excited modes, a resonance may occur. This resonance will cause the frequency difference of two excited modes to lock at $\nu_s/2$ or $\nu_s$.

%\com{Write some more theory about NS and BH blah blah}

Now for instance a system with $l$ degrees-of-freedom has $l$ linear natural frequencies and modes denoted by $\nu_1, \cdots, \nu_l$ with all being real and non-zero. Then if the frequencies are commensurable or nearly commensurable, the system exhibits a strong frequency coupling, giving rise to internal resonance \citep{nayfeh1979nonlinear}. In addition, if there exists a harmonic external excitation frequency $\nu_s$, then the frequencies might have a commensurable relationship exhibiting the resonance as
\eq{nu_s_Sigma}{
p\nu_s = \Sigma_{i = 1}^l b_i\nu_i,
}
where $p$ and $b_i$ are integers such that
\eq{}{
p+\Sigma_{i = 1}^n |b_i| = k +1,
}
where $k$ is the order of non-linearity. 

As described in section \ref{sec:mod_ep}, we know that the fundamental frequencies of the accretion system will be the modified epicyclic frequencies for the fluid system as given by equations \Eref{mod_QPO_main_1} and \Eref{mod_QPO_main_2}. Therefore, similar to \citealt{Mukhopadhyay_2009}, we rewrite the equation \Eref{nu_s_Sigma} for an accretion disk with the 2 degrees-of-freedom with corresponding frequencies $\nu_1$ and $\nu_2$ as
\eq{nu_s_sigma_nm}{
(n-m+1)\nu_s = b_1\nu_1 + b_2\nu_2.
}
%where $\nu_1$ and $\nu_2$ are the modified QPO frequencies that we obtain in equation \Eref{mod_QPO_main_1} and \Eref{mod_QPO_main_2}. 

\subsection{Nonlinear regime}
Here we take $b_1 = - b_2 = 2$, which leads to 
\eq{QPO_1}{
\frac{\nu_s}{2}(n - m + 1) = \nu_1 - \nu_2,
}
with $m$ and $n$ being integers. In the disk around a neutron star, at an appropriate radius where $\nu_1 - \nu_2 \approx 0$ the resonance is supposed to take place, which from equation \Eref{QPO_1} gives $n = m-1$. 
%And the difference in high and low frequencies $\Delta\nu\approx \nu_s/2$ leads to a mode locking in the nonlinear regime. 
Now we propose the higher and lower QPO frequencies of a pair to be, respectively,
\begin{eqnarray}
    \nu_h = \nu_1 + \frac{m+1}{2}\nu_s,\,\,\
    \nu_l = \nu_2 + \frac{m}{2}\nu_s.
    \label{nuhl}
\end{eqnarray}

After rearranging the terms of equation \Eref{QPO_1} we obtain the relation as 
\eq{}{
\left(\nu_1 + \frac{m+1}{2}\nu_s\right) - \left(\nu_2 + \frac{m}{2}\nu_s\right) = \frac{\nu_s}{2}.
}
From our proposed QPO frequencies and resonance condition, we see that the condition for $\Delta\nu$ is satisfied as
\eq{}{
\Delta\nu = \nu_h - \nu_l \approx\frac{\nu_s}{2}.
}
Fig. \ref{fig:Freq_dist} (left panel) shows the formation of the above resonance and QPOs at around $r=9.49GM/c^2$. At this radius $\nu_1=\nu_2$ and $\Delta\nu\sim\nu_s/2$. The neutron star mentioned there is reported in the following section.                                                                                

The observed phenomenon of the frequency difference between twin kHz QPOs remaining close to the half of the stellar spin frequency (or the stellar spin frequency itself, as explained below) in many neutron star systems, even if the QPO frequencies vary over time, can be explained by the above model. Even though the excitation order $m$ changes, the difference in the high and low frequencies remains the same.

\subsection{Linear regime}
\begin{figure*}
\centering
   \includegraphics[width=.4\textwidth]{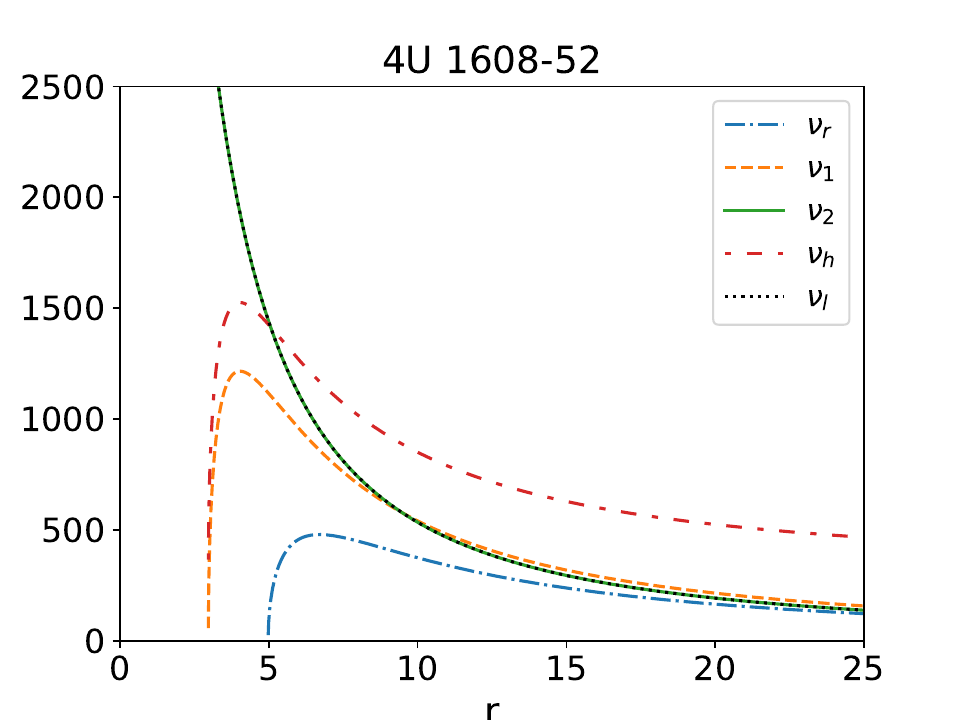}
   \includegraphics[width=.4\textwidth]{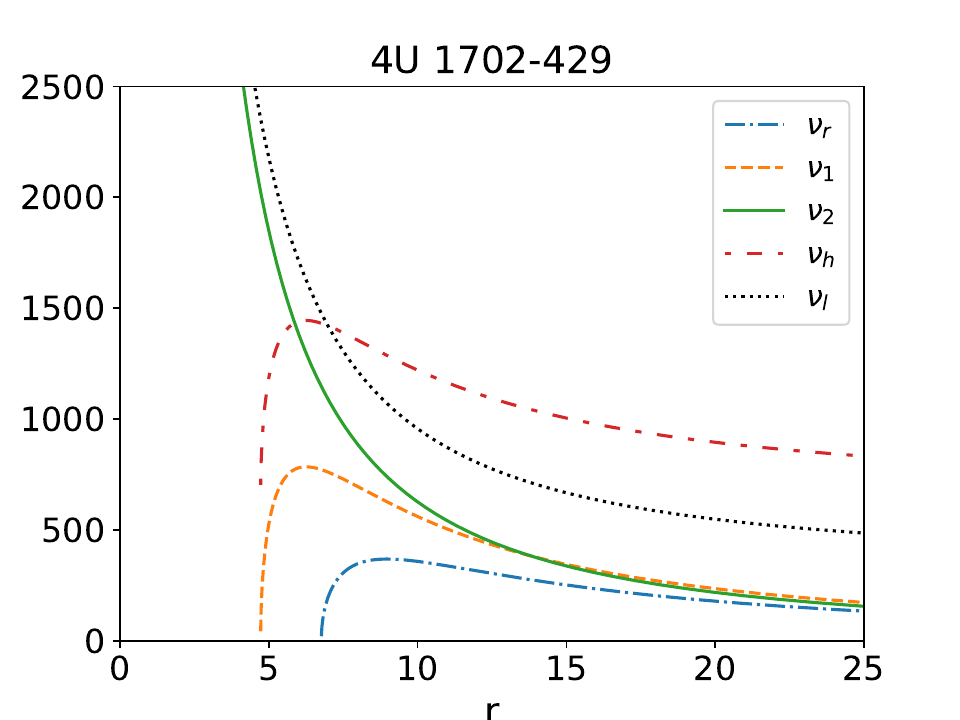}
   \hfill
   \caption{ Variation of different frequencies (in Hz) as a function of distance from the central object in units of gravitational radius in the accretion disk, where (a) 4U 1608-52 is a fast rotator with the resonance radius $r=9.49GM/c^2$, and (b) 4U 1702-429 is a slow rotator with the resonance radius $r=13.8GM/c^2$. The value of $m$ is chosen to be 0 and 1 respectively so that they lead to the reasonable radii, as given in Table \ref{tab:NS_param_FR}. 
   } 
   \label{fig:Freq_dist}
\end{figure*}

However, for a black hole due to the absence of a magnetosphere and a weakly magnetized neutron star, the coupling between modes may not be nonlinear and resonance locking may occur in the linear regime when the condition $\nu_h - \nu_l$ being approximately equal to $\nu_s$ is met. For this condition, we just have to take the first order excitation in equation \Eref{nu_s_sigma_nm} and put $b_1 = - b_2 = 1$, and following the same strategy, as in nonlinear regime, the QPO frequencies take the form 

\begin{eqnarray}
    \nu_h &=& \nu_1 + (m+1)\nu_s,\label{eq:QPO_BH_1}\\
\nu_l &=& \nu_2 + m\nu_s.\label{eq:QPO_BH_2}
\end{eqnarray}
It is evident from equations \Eref{QPO_BH_1} and \Eref{QPO_BH_2} that the condition for resonance $\Delta\nu\approx\nu_s$ is satisfied. 
%This is the case for some neutron stars as well, as maintained above, which have a weak magnetic field.
Fig. \ref{fig:Freq_dist} (right panel) shows that at $r=13.8GM/c^2$ the linear resonance forms with $\nu_1=\nu_2$ and $\Delta\nu\sim\nu_s$. The mentioned neutron star therein is reported in the following section. 

\section{Basic characteristics of black holes and neutron stars}

As mentioned above, there is a subtle difference  
in the resonance conditions for the formation of QPOs around black holes and neutron stars. This may be attributed to the disparities in their environment. In the case of a magnetized neutron star particularly, the oscillations of the disk are directly influenced by the rotating magnetosphere, which imprints the spin frequency of the star leading to a resonance. However, for a black hole, the energy and angular momentum may be transferred to the surrounding region through the magnetic field lines, which however may not be as strong as of neutron stars, particularly for soft states \citep{BZ77}. This is expected to lead to a strong resonance, to be driven by the disturbances at the spin frequency of the black hole, although the exact mechanism is yet to be determined. Additionally, it should be noted that while the presence of a hard surface of a neutron star leads to a strict definition of $\nu_s$ corresponding to its boundary layer related to the spin parameter $a$, there is no such definition for a black hole. Nevertheless, here we assume that the spin of the black is experienced by the accreting fluid through the frame dragging frequency.

The spin frequency of a neutron star can be asserted from observed data. However, in computing the QPO frequencies from our model, we also need to determine the dimensionless spin parameter $a$. Assuming the neutron star to be almost spherically symmetric with the equatorial radius $R$, spin frequency $\nu_s$, mass $M$, the radius of gyration $R_G$, the moment of inertia and the spin parameter, respectively, can be defined as
\eq{NS_a}{
I = M R_G^2, \hspace{0.5cm} a = \frac{I \Omega_s}{\frac{GM^2}{c}},
}
where $\Omega_s = 2\pi\nu_s$. However for a black hole, mass $M$ and $a$ are the most natural parameters that we supply as inputs. The corresponding frame dragging angular frequency on the spacetime around the black hole is then given by
\eq{}{
\Omega_{BH} = -\frac{g_{\phi t}}{g_{\phi\phi}} = \frac{2a}{r^3 + ra^2 + 2a^2},
}   
and, thus, for a black hole, we define the spin frequency as 
\eq{}{\nu_{BH}=\nu_s = \frac{\Omega_{BH}}{2\pi}\frac{c^3}{GM},}
which is actually the imprint of spin frequency at a given radius away from the black hole. 
Therefore, by supplying the spin parameter $a$, mass $M$ and the radius $r$, the effective spin frequency can be determined. 

\section{Properties of neutron star QPO}
It has been observed that QPO frequencies tend to vary over time, with the frequency difference between them in a pair decreasing slightly as the lower QPO frequency increases. As mentioned above, for fast rotators, the frequency difference is approximately half of the spin frequency, whereas for slow rotators, it is of the same order as the spin frequency. Our model successfully explains these observed properties as demonstrated below. To obtain our results, we explore a suitable range of mass of the neutron star $M$ and the spin parameter $a$. For $\nu_l$ and $\nu_h$, we use the Kerr metric only, as given by equations (\ref{nuphi}), (\ref{nur}), (\ref{nuz}), (\ref{eq:mod_QPO_main_1}), (\ref{eq:mod_QPO_main_2}). However, strictly speaking the Kerr metric is for black holes. Nevertheless, for the present purpose, it does not matter practically as we are interested outside of a neutron star and also not to close to it. Now, using the grid search algorithm, we find the parameters that fit the observational data the best. 
Basically we try to find out the parameters that make reduced $\chi^2=\sum_i(O_i-C_i)^2/\sigma^2_i D$ close to unity, here $O_i$ and $C_i$ are observed and our model computed values corresponding to the $i^{\rm th}$ observation, respectively, with $\sigma_i$ as the variance of measurement error and $D$ is the degrees of freedom (i.e. the number of observed data points minus the number of fitted parameters). Our aim should be to align $\chi^2$ as close to unity as possible. A suitable function to gauge the closeness of $\chi^2\rightarrow 1$ would be $f(\chi^2) = |1 - \chi^2|$. This function reaches its minimum value of zero when $\chi^2\rightarrow 1$, thus implying a good fit. Therefore, we plan to find the parameters that minimize $f$ (ideally $\rightarrow 0$), thereby ensuring that $\chi^2$ is as near to one as possible, indicative of an optimal fit.

\begin{table*}
    \centering
    \begin{tabular}{lccccccc}
        \hline
        \hline
        Fast Rotator &  $\nu_s$\ (Hz) & $M \ (M_\odot)$ & $R$ (km)  & $m$& $|1-\chi^2|$\\
        \hline
        \hline
                 KS 1731-260 &  524 & 1.51  & 10.08 - 12.04 & 1 & -NA-\\

          KS 1731-260 &  524 & 1.2  & 10.32 - 12.33 & 0 & -NA-\\

        KS 1731-260 &  524 & 1.07  & 10.99 - 13.14 & -1 & -NA-\\
        %KS 1731-260 &  524 & 0.96 & -0.24 & 7.87 - 9.41 & 1 & -NA-\\
        KS 1731-260 &  524 & 1.07 & 8.31 - 9.93 & 0 & -NA-\\
        
        %KS 1731-260 &  524 & 0.9 & 0.3 & 8.52 - 10.18 & -1 & -NA-\\
        \hline
        \hline
        \multirow{2}{*}{4U 1636-53}&\multirow{2}{*}{581.75} & \multirow{2}{*}{1.5} & \multirow{2}{*}{14.52 - 17.35} & -2 & \multirow{2}{*}{0.689}\\&& & &1\\
        \hline
        \multirow{2}{*}{4U 1636-53}&\multirow{2}{*}{581.75} & \multirow{2}{*}{1.65}  & \multirow{2}{*}{15.61 - 18.66} & -2 & \multirow{2}{*}{0.797}\\&& & & 0\\
        \hline
        \multirow{2}{*}{4U 1636-53}&\multirow{2}{*}{581.75} & \multirow{2}{*}{1.23} & \multirow{2}{*}{10.21 - 12.21} & -1 & \multirow{2}{*}{0.966}\\& & & &1\\
 
        \hline
        \hline
        
         4U 1608-52 &  619 & 1.87  & 13.69 - 16.36 & 1 & 2.54\\
         4U 1608-52 &  619 & 1.86 & 11.27 - 13.47 & 0 & 2.47\\
         4U 1608-52 &  619 & 1.6 & 13.08 - 15.64 & -1 & 2.67\\
        \hline
        \hline
        
        Slow Rotator &  $\nu_s$\ (Hz) & $M \ (M_\odot)$ & $R$ (km)  & $m$\\
        \hline
        4U 1702-429 &  330 & 1.67 & 9.25 - 11.06 & 1 & $\sim$0.00\\
        4U 1702-429 &  330 & 0.85  & 7.62 - 9.11 & 0 & $\sim$0.00\\
        4U 1702-429 &  330 & 1.2 & 11.98 - 14.32 & -2 & $\sim$0.00\\
        4U 1702-429 &  330 & 0.97 & 10.77 - 12.87 & -3 & $\sim$0.00\\
        
        \hline
        \hline

        \multirow{2}{*}{4U 1728-34}&\multirow{2}{*}{364.23} & \multirow{2}{*}{1.99} & \multirow{2}{*}{15.2 - 18.16} & -2 & \multirow{2}{*}{3.046}\\& & & &-1\\
        \hline
        
        \multirow{2}{*}{4U 1728-34}&\multirow{2}{*}{364.23} & \multirow{2}{*}{1.26} & \multirow{2}{*}{11.58 - 13.84} & -3 & \multirow{2}{*}{3.137} \\&& & &-2\\
        \hline

        \multirow{2}{*}{4U 1728-34}&\multirow{2}{*}{364.23} & \multirow{2}{*}{0.85} & \multirow{2}{*}{8.88 - 10.62} & -4 & \multirow{2}{*}{3.226} \\& & & &-3\\

        \hline
        \hline
         &  Estimated $\nu_s$\ (Hz) &  & & & \\
        \hline
        \hline
        Sco X-1 &  300 & 1.4 & 22.5 - 26.9 & 1 & 0.35\\
        Sco X-1 &  600 & 1.4 & 15.91 - 19.08 & 1 & 0.35\\
            Sco X-1 &  662 & 1.44 & 10.21 - 12.2 & -1 & 0.75\\
        Sco X-1 &  602 & 1.89 & 11.71 - 14 & 1 & 0.35\\
         \hline
        \hline
        \end{tabular}
    \caption{Physical parameters for neutron stars.}
    \label{tab:NS_param_FR}
\end{table*}

\subsection{Fast Rotators}\label{sec:Fast_Rotator}
\begin{figure*}
\centering
   \label{KS1731-260} \includegraphics[width=.33\textwidth]{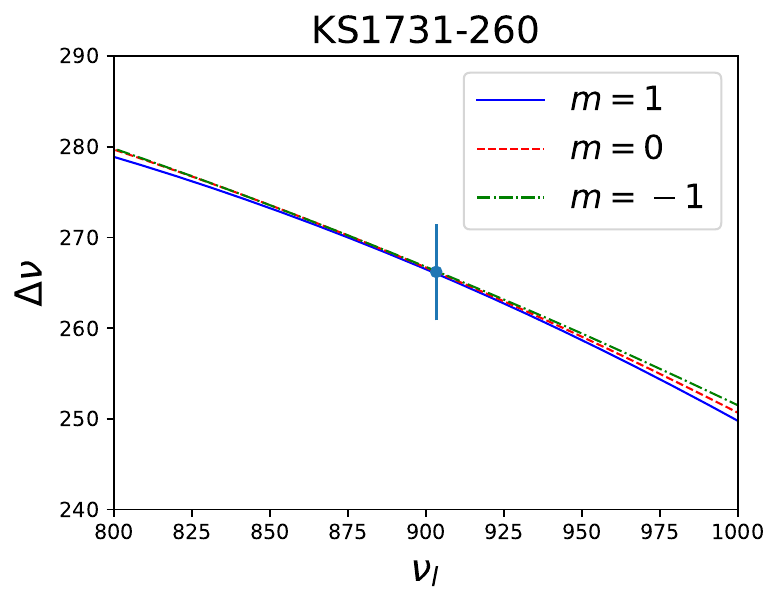}\hfill
   \label{4U 1636-53} \includegraphics[width=.33\textwidth]{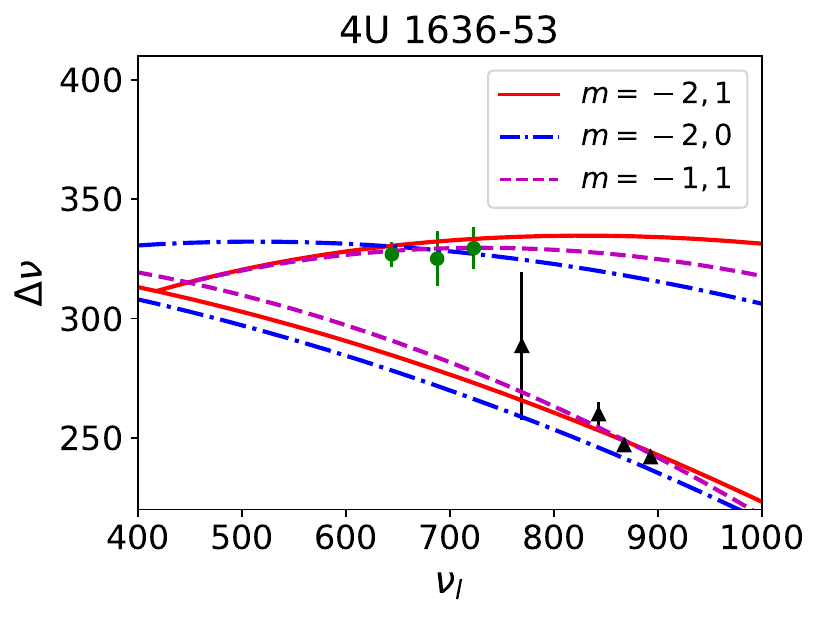}\hfill
   \label{4U 1608-52}\includegraphics[width=.33\textwidth]{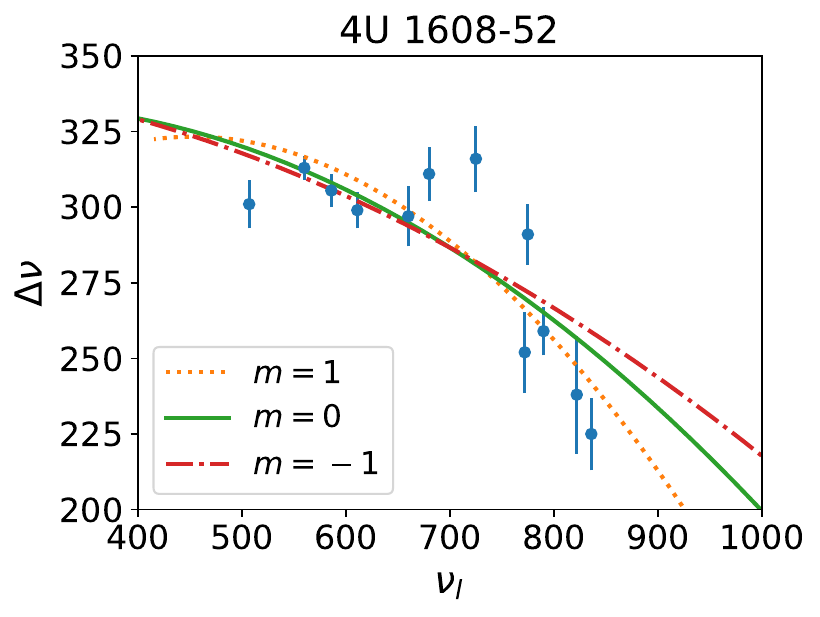}
	\caption{Variation of QPO frequency difference with lower QPO frequency for fast rotators: (a) KS 1731-260, (b) 4U 1636-53, and (c) 4U 1608-52. The points with the error-bar are observed data and the lines correspond to the model fitting. For (b) the data have been fitted for two modes shown by green circles and black triangles. See Table \ref{tab:NS_param_FR} for other details for each source including $\chi^2$.} \label{Fast Rotator}
\end{figure*}

We analyze the QPOs of some of the fast rotating neutron stars whose spin frequencies are known: KS 1731-260 \citep{Smith_1997}, 4U 1636-53 \citep{Jonker_2002} and 4U 1608-52 \citep{Méndez_1998}. We vary the mass $M$ and the spin parameter $a$ in appropriate and/or admissible ranges and find their suitable values giving the best fit for the observed QPOs based on our model. 

We know that for a solid sphere we have $R_G^2 = 2R^2/5$ and for a hollow sphere $R_G^2 = 2R^2/3$.  It is known that neutron stars are not perfect solid bodies and are expected to be deformed from a perfect spherical to an ellipsoidal shape at very fast rotation rates. Hence, in our calculation, we consider a range: $0.35\le (R_G/R)^2\le 0.5$ for most cases, as suggested in previous studies \citep{Bejger_2002,1994ApJ, Mukhopadhyay_2009}. Hence, from equation \Eref{NS_a} we have the radius of the neutron star in the range
\eq{R_NS}{\frac{R_G}{\sqrt{0.35}}\le R\le\frac{R_G}{\sqrt{0.5}},}
where $R_G = \sqrt{{GMa}/{2\pi\nu_sc}}$. Table \ref{tab:NS_param_FR} shows the estimated parameters that fit best the observed QPOs based on our model. 

%\com{Caution that many will give $chi^2$ to be zero hence sorting based on this will not work. \\ (Sol): The chi2 appeared to be zero uptill two decimal places, it will not be actually zero, hence our analysis is correct as the sorting will work}

KS 1731-260, to date, has exhibited only one pair of QPO frequencies, established by our model and is depicted in Fig. \ref{Fast Rotator}. However, due to its single data point, many parameters can reproduce the observed QPO frequencies within the error bound.  Therefore, we only consider the non-linear regime for $m = -1, 0$ and 1. We also find that considering other values of $m$ or linear regime gives a radius  $R > 12.5$ km, which is strongly restricted by \cite{Ozel_2012}. Using our model, we obtain $M = 1.51M_\odot$ with a radius between $10.08$ km and $12.04$ km; $M = 1.2M_\odot$ with a radius in $10.32$ km and $12.33$ km; and $M = 1.07M_\odot$  with a radius in $10.99$ km and $13.14$ km, for $m = -1, 0$, and 1, respectively.
However, if we allow lower mass and radius, e.g. like a strange star, our model also produces the result (see, e.g., \citealt{BMApJL2003}), shown one
such a case in Table \ref{Fast Rotator}. 

For 4U 1636-53  and 4U 1608-52, however, several twin peak QPOs have been observed varying with time. Although the high and low QPO frequencies vary, their difference remains almost constant with a slightly decreasing trend with the increasing lower QPO frequency. This trend has been successfully reproduced and the observed QPOs have been fitted very well by our model.

For 4U 1636-53, from the data it is evident that the source is in non-linear regime like other fast rotators. But $\Delta\nu$, though remains constant for lower $\nu_l$, decreases little more rapidly with the increasing $\nu_l$, compared to other cases. Hence this gives a rather poor fitting if we consider activation of only one mode $m$,
as shown in Fig. \ref{Fast Rotator}. However, it could be possible that two harmonics (or modes), say $m_1$ and $m_2$, are activated at various times of observations, and not considering both the modes appropriately in the analysis results in a poor fit. To address this issue, we vary the mass $M$ and spin parameter $a$ for two $m$-s: $m_1$ and $m_2$. This gives two sets of curves for a given $M$ and $a$ in the nonlinear regime, yielding two sets of parameters (with two modes or $m$ values), as shown in Fig. \ref{Fast Rotator}. One set results in a mass of $M = 1.5M_\odot$, corresponding to a neutron star radius range of $14.52 \text{ km }\le R \le 17.35 \text{ km}$, a bit large, for $m = -2, -1$. The other set results in a mass of $M = 1.65M_\odot$, corresponding to a neutron star radius range of $15.61 \text{ km }\le R \le 18.66 \text{ km}$, even larger, for $m = -2, 0$. However, for $m=-1, 1$, with a slightly lower mass $M=1.23M_\odot$, the radius range turns out to be $10.21\text{ km }\le R \le 12.21 \text{ km}$. See Table \ref{tab:NS_param_FR}.

The mass of 4U 1608-52 was already estimated to be 
$M = 1.74 \pm 0.14 M_\odot$ \citep{G_ver_2010}. Thus the mass parameter has been varied in this range yielding three sets of parameters. The observed QPO frequencies are best reproduced for  mass $M  = 1.87M_\odot$, with radius of neutron star in range 13.69 km to 16.36 km; $M = 1.86M_\odot$ corresponding to radius in range 11.27 km to 13.47 km, and finally for mass $M = 1.6M_\odot$ for the radius in 13.08 km to 15.64 km.  

%From our model we find the mass to be $M = 1.59M_\odot$  which, from equation \Eref{R_NS}, gives the radius to be in the range of $7.84 \text{ km }\le R \le 9.38 \text{ km }$, which matches with earlier result \citep{G_ver_2010}. The fitting has been shown in Fig \ref{Fast Rotator}. 

\subsection{Slow Rotators}

\begin{figure*}
\centering
   \label{4U1702-429} \includegraphics[width=.33\textwidth]{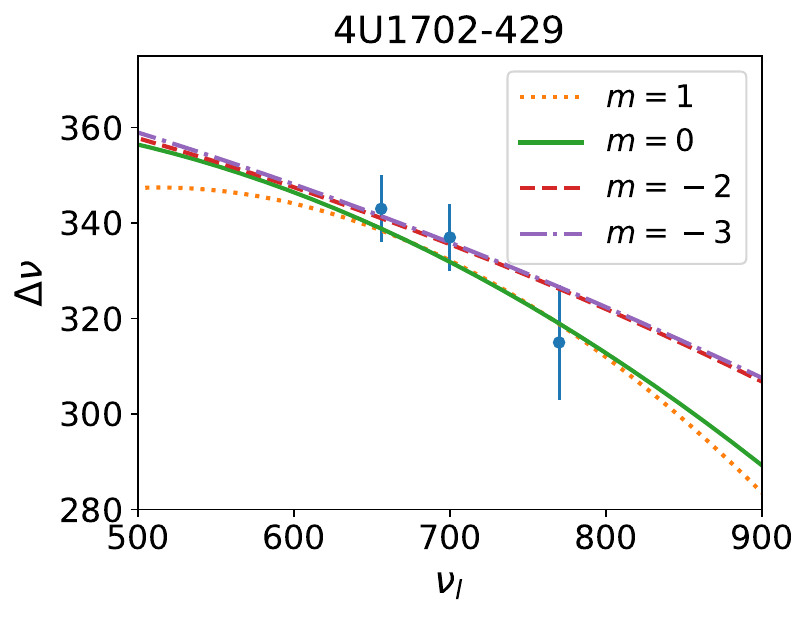}\hfill
   \label{4U1728-34} \includegraphics[width=.33\textwidth]{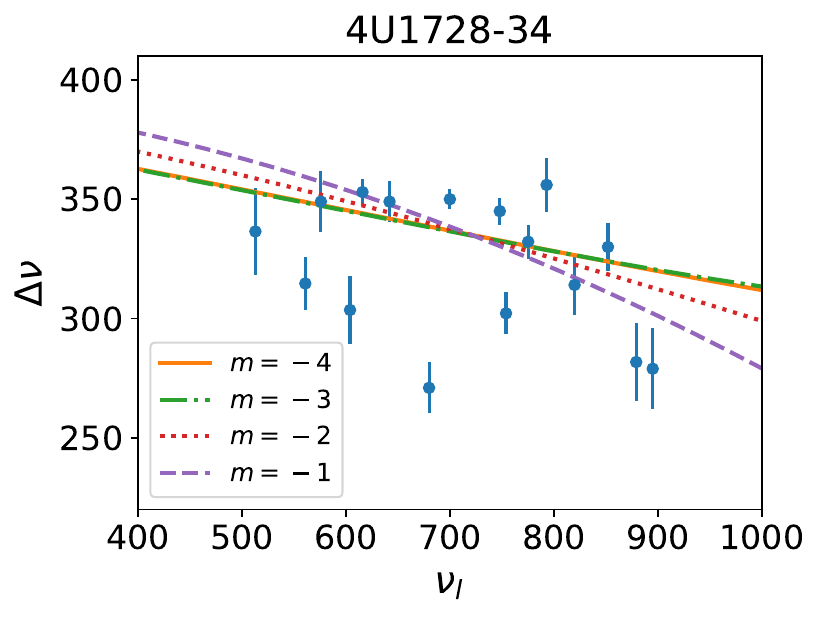}\hfill
   \label{Sco X-1}\includegraphics[width=.33\textwidth]{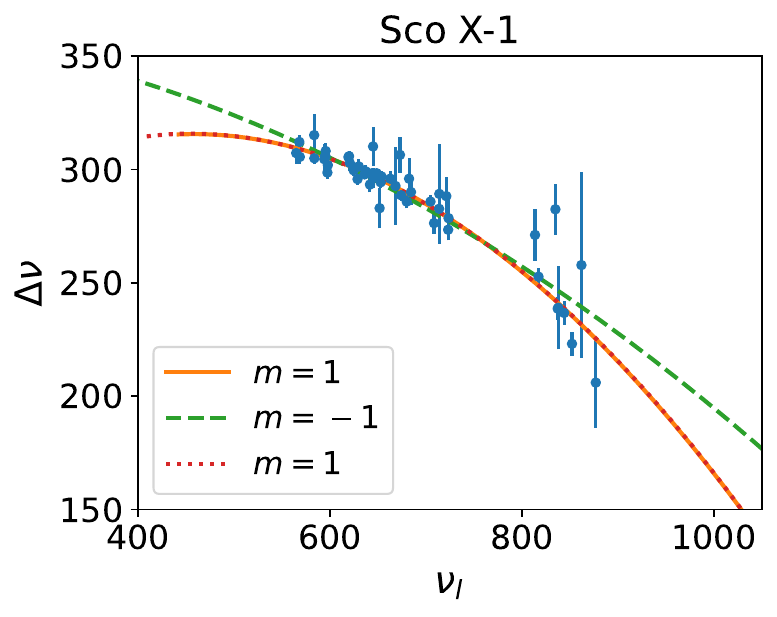}
	\caption{Same as Fig. \ref{Fast Rotator}, except for two slow rotators: (a) 4U 1702-429, and (b) 4U 1728-34; and (c) Sco X-1 whose spin frequency is not confirmed yet. See Table \ref{tab:NS_param_FR} for other details for each source including $\chi^2$.} \label{Slow Rotator}
\end{figure*}

\begin{figure}
\centering
    \includegraphics[width=.36\textwidth]{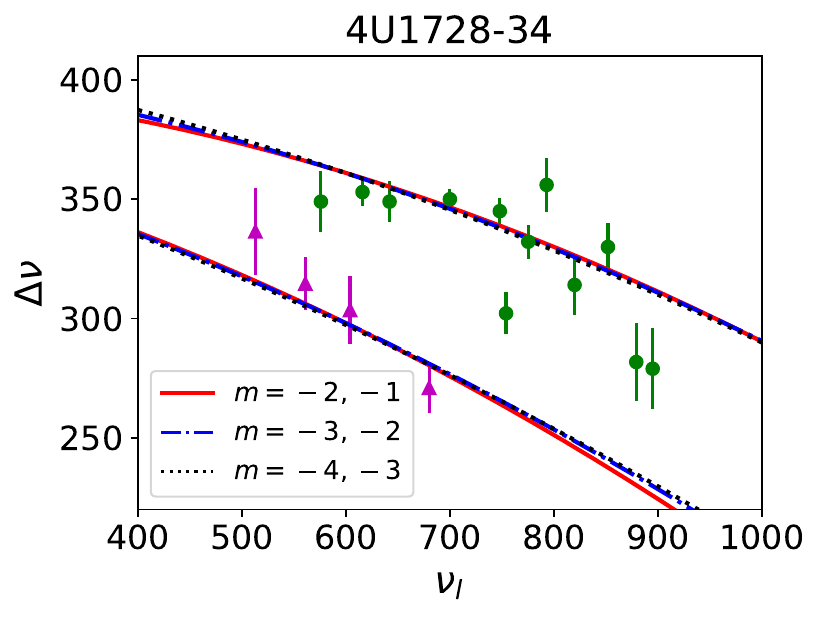}
	\caption{Same as Fig. \ref{Fast Rotator}, except for  4U 1728-34. The observed QPO frequencies have been fitted for two modes shown by green circle and magenta triangle for a fixed mass and spin parameter. See Table \ref{tab:NS_param_FR} for other details including $\chi^2$.} \label{4U1728-34ver2}
\end{figure}
%The spin of 4U0614+091 have been estimated using during a thermonuclear X-ray burst to be $\approx415$ Hz \citep{Strohmayer_2008} \com{make changes}
The slow rotating neutron stars with known spin frequencies
are: 4U 1702-429 \citep{Markwardt_1999}, 4U 1728-34
\citep{van_Straaten_2002,Méndez_1999}. 
%Similar to the previous section here also we vary the mass $M$ and the spin parameter $a$. Since we don't want the radius to get to very large values hence we consider the mass $M<1.4M_\odot$. 
It is evident from the data that the sources are in the linear regime. By varying the mass and the spin parameter we obtain best fitting parameters. For 4U 1702-429, we find that mass $M = 1.67M_\odot$ gives the radius to be in 9.25 km to 11.06 km. The other solutions that also fit well with the data correspond to mass $M<1.4M_\odot$; with the mass being $M = 0.85M_\odot, 1.2M_\odot$ and $0.97M_\odot$. The corresponding $m$ and the radius are given in Table \ref{tab:NS_param_FR} for both the stars and the fitting is shown in Fig. \ref{Slow Rotator}.

However, the relationship between the lower kHz QPO frequency ($\nu_l$) and the difference of frequencies ($\Delta\nu$) for 4U 1728-34 does not follow the expected trend where $\Delta \nu$ seems to have a bimodal trend with $\nu_l$. This causes a rather poor fitting of the data by our model for a fixed $m$, as shown in Fig. \ref{Slow Rotator}. As discussed in section \ref{sec:Fast_Rotator} for the case of 4U 1636-53, it is possible that two harmonics or modes, $m_1$ and $m_2$, are activated at different times during observations, and not considering both the modes adequately in the analysis results in a poor fit. Hence similar to the case of 4U 1636-53, we vary the mass $M$ and spin parameter $a$ for two $m$-s. 
This gives two curves simultaneously for a given $M$ and $a$ which are fitted with the observed data as shown in Fig. \ref{4U1728-34ver2}. For this, the estimated mass comes out to be $M = 1.99M_\odot$ with the radius in the range 15.2 km to 18.16 km, $M = 1.26M_\odot$ corresponding to radius in 11.58 km to 13.84 km, and $M = 0.85M_\odot$ has the radius in 8.88 km to 10.62 km, as mentioned in Table \ref{tab:NS_param_FR}. But following \cite{Mukhopadhyay_2009},  we can stick to mass $M<1.4M_\odot$.

%This gives two curves simultaneously for a given $M$ and $a$ which are fitted with the observed data as shown in Fig. \ref{4U1728-34ver2}. For this, the estimated mass comes out to be $M = 1.38M_\odot$ with the radius range being 17.6 km to 21.03 km as mentioned in Table \ref{tab:NS_param_FR}, if we consider the neutron star to be of spherical shape the corresponding the radius range becomes 15.24 km to 19.67 km. \\

\subsection{Estimate of Spin Frequency for Sco X-1}
Sco X-1 is an X-ray binary system with a low-mass companion star of mass approximately $0.42M_\odot$ and a neutron star of mass around $1.4M_\odot$, as reported by \cite{Steeghs_2002_1,PhysRevD.92.023006}. Although the spin frequency of Sco X-1 remains unknown, this source has been observed to exhibit a pair of kHz QPO frequencies with $\Delta\nu$ in a range of $\sim 225 - 310$ Hz \citep{10.1046/j.1365-8711.2000.03788.x}.
Here we vary the mass from $1.35M_\odot$ to $1.45M_\odot$ so that it remains close to the $1.4M_\odot$ value and we vary the spin frequency $\nu_s$ from 200 Hz to 800 Hz. Now if we consider the nonlinear mode locking, then the best fit $\nu_s$ turns out to be at 600 Hz for the radius range 15.91 to 19.08 km; and 662 Hz for the radius range 10.21 to 12.2 km. However, the former seems to suggest a very high radius, hence is uncertain. In the linear regime, i.e. considering it to be a slow rotator, we find that the 
%for it to be a slow rotator. We thence fit the variation of observed lower QPO frequency with the frequency difference and see what parameters fit the observed data the best in the linear regime. Since the spin frequency $\nu_s$ of this source is unknown, we take this as a varying parameter along with the spin parameter $a$, while the mass is kept fixed at $M = 1.4M_\odot$. We find that the 
spin frequency of $\nu_s = 300$ Hz gives an excellent fit with the observed data, as shown in Fig. \ref{Slow Rotator}. However, it seems to be ruled out due to quite a large possible radius given in Table \ref{tab:NS_param_FR}. Although we have varied the mass in a range around $1.4M_\odot$, we still have obtained the best fit to be at $M=1.4-1.44M_\odot$ with all the corresponding parameters as mentioned in Table \ref{tab:NS_param_FR}. 
If the mass is relaxed to a higher value, e.g., $M=1.89M_\odot$, then the best-fit radius range in the nonlinear regime turns out to be 11.71 to 14 km with $\nu_s=602$ Hz. 

\section{Black hole QPOs}
\begin{table*}
    \centering
    \begin{tabular}{lcccccc}
    \hline
    \hline
        Black hole &  $M\ (M_\odot)$ & Estimated $a$ & $\nu_h$\ (Hz) & $\nu_l$\ (Hz) & $m$ \\
        \hline
        \hline
	    GRO J1655-40 &  $7.02^{+0.22}_{-0.2}$ & $0.91\mp0.02$ & \textbf{450} & \textbf{300} & -1\\
        GRO J1655-40 &
	    \edits{ $5.4_{-0.3}^{+0.08}$} & \edits{$0.8_{+0.06}^{-0.19}$}  & \textbf{450} & \textbf{300} & \edits{-2}\\
        %$5.31^{+0.02}_{-0.04}$ & $0.82\mp0.01$ & 450 & 300 & -2\\
        \hline
        \hline
	    XTE J1550-564 & 10.92 - 11.07 &  $0.57\pm0.01$ & \textbf{276} & \textbf{184} & -1\\
	    XTE J1550-564 & 10.26 - 11.58 &  0.99 - 0.89 & \textbf{276} & \textbf{184} & -1\\
        \hline
        \hline
	    H1743-322 & \edits{9.97 - 10.91} & \edits{0.61 - 0.7} &  \textbf{240} & \textbf{163} & \edits{-2} \\
	    H1743-322 & \edits{9.53 - 9.61} & \edits{$0.6\pm0.01$} &  \textbf{240} & \textbf{163} & \edits{-2} \\
	    H1743-322 & \edits{9.25 - 10.86} & \edits{0.87 - 0.71} &  \textbf{240} & \textbf{163} & \edits{-2} \\
        \hline
        \hline

	    XTE J1859+226  & 5.9 - 9.69 & 0.57 - 0.61 & \textbf{227.5} & \textbf{128.6} & -2 \\
        \hline
        \hline
	    IGR J17091-3624 & 9.9 - 10.02 & $0.59\pm0.01$ &  164 & \textbf{66} & -2\\
	    IGR J17091-3624 & 11.04 - 15.46 & 0.6 - 0.67 &  164 & \textbf{66} & -2\\
	    IGR J17091-3624 & 8.89 - 10.63 & 0.66 - 0.99 &  164 & \textbf{66} & -3\\
        \hline
        \hline
        
	    \multirow{2}{*}{GRS 1915+105} & \multirow{2}{*}{ $12.4^{+2.0}_{-0.75}$}& \multirow{2}{*}{ $0.94_{+0.05}^{-0.14}$ } & \textbf{168} & 113 &-2\\
	    &&& \textbf{69} & \textbf{41} & -6\\
        %&&& \color{blue}\textbf{$\sim$71.5} & \color{blue}{$\sim$ 60} & \color{blue}-2\\
        %  &&& \color{blue}{41} & \color{blue}{34} & \color{blue}-2\\

        % GRS 1915+105 &  $12.4^{+2.0}_{-0.75}$ & $0.94_{+0.05}^{-0.14}$ & \textbf{168} & 113 & -2\\
        % \hline
        % GRS 1915+105 &  $12.4^{+2.0}_{-0.75}$ & $0.94_{+0.05}^{-0.14}$ & \textbf{$\sim$71.5} & {$\sim$ 60} & -2\\
        % \hline
        % GRS 1915+105 &  $12.4^{+2.0}_{-0.75}$ & $0.94_{+0.05}^{-0.14}$ & {$\sim$41.1} & {$\sim$34.1} & -2\\
        \hline
        \hline
        XTE J1752-223 & $10^{+1.63}_{-1.45}$ & $0.62^{+0.3}_{-0.2}$ & 442.91 & 114.87 & -1\\
        \hline
        \hline
        \multirow{2}{*}{GX 339-4}&\multirow{2}{*}{5.3 - 6.17} & \multirow{2}{*}{0.65 - 0.99} & 278.67 & 127.7 & -3\\
        & & & 208.44 & 105.35 & -4 \\
        \hline
        \multirow{2}{*}{GX 339-4}&\multirow{2}{*}{6.13 - 6.17} & \multirow{2}{*}{$>0.97$} & 278.67 & 127.7 & -3\\
        & & & 208.44 & 105.35 & -4 \\
        \hline
        \hline
    \end{tabular}
    \caption{Physical parameters for black boles. The boldfaced frequencies are the ones without dispute as of now.}
    \label{tab:BH_param}
\end{table*}
Several black holes with twin HF QPOs have their masses determined through independent observations, such as GRO J1655-40 \citep{Orosz_1997, 10.1046/j.1365-8711.1999.02481.x}, XTE J1550-564 \citep{Orosz_2002}, GRS 1915+105 \citep{Greiner2001, Reid_2014}, H1743-322 \citep{Miller_2006} and IGR J17091-3624 \citep{Altamirano_2012}. However, the spin parameter of these black holes remains uncertain. To estimate the spin of these black holes, we fit the observed QPOs to our model and determine the spin parameter that provides the best fit, as summarized in Table \ref{tab:BH_param}. We search for the optimal value of $a$ by varying the mass of the black hole within its error bar. 

The black hole mass of GRO J1655-40 is still a matter of debate, with some groups estimating it to be  
$M = 7.02 \pm0.22 M_\odot$ \citep{Orosz_1997}, 
\edits{while others suggest $M = 5.4\pm0.3$ \citep{Beer_Podsiadlowski_2002} and
$5.31 \pm 0.07 M_\odot$ \citep{10.1093/mnras/stt2068}}.
To account for this uncertainty, we consider all the suggested mass ranges (mass values with the error bars) and estimate the spin parameter. The two HF QPOs now known in this source occur at frequencies of 300 Hz and 450 Hz \citep{10.1093/mnras/stt2068}. We find that not the entire mass ranges are able to reproduce the observed QPOs. We find that for $M$ to be in $6.82 - 7.24 M_\odot$ and \edits{ $5.1 - 5.48 M_\odot$}, the observed QPOs are reproduced when the spin parameter $a$ lies in $0.93 - 0.89$ for $m = -1$ and \edits{$0.86- 0.61$} for $m = -2$, respectively. 

The mass of the black hole XTE J1550-564 has been taken to be $M = 9.68 - 11.58 M_\odot $ \citep{Orosz_2002}, which shows HF QPOs at frequencies 276 Hz and 184 Hz \citep{Remillard_2002}. We find that for the mass $M$ to be in the range of $10.92 M_\odot$ to $11.07 M_\odot$ and $m = -1$, the spin parameter is to be $a = 0.57 \pm 0.01$ in order to obtain observed QPOs by our model. Also for $M$ to be in $10.26 - 11.58 M_\odot$, $a$ is estimated to be in $0.99 - 0.89$ for $m = -1$. However $a = 0.57$ is in very close agreement with the spin parameter obtained by Fe line fitting as by \cite{Steiner_2011}, This leads to a better constraint on the mass of the black hole in XTE J1550-564.

\edits{
The source H1743-322 has been observed with HF QPO peaks at 240 Hz and 163 Hz  \citep{2005ApJ...623..383H}}. The mass of this black hole is estimated to be in the range of $9.25\text{ to }12.86M_\odot$ \citep{Molla:2016mip}. It has been shown by \cite{Steiner_2011_2} that the spin parameter $a$ of this black hole lies in the range of -0.3 to 0.7 and the extreme cases of $a \sim 0.9$ can be ruled out with high confidence. Considering this and after fitting our model best with the observed QPOs, we obtain the mass of the black hole to be in the range of \edits{$9.53\text{ to }9.61M_\odot$  for $a = 0.6\pm0.01$, and in $9.97\text{ to }10.91M_\odot$ for $a = 0.61\text{ to }0.7$, and in range of 9.25 to 10.86 $M_\odot$ for $a = $ 0.87 to 0.71.
}

Based on the optical observations in 2017, the black hole XTE J1859+226 during quiescence is estimated to have a mass of $M = 7.8 \pm1.9 M_\odot$ \citep{10.1093/mnras/stac2719}. In addition, a pair of QPOs with frequencies of 128.6 and 227.5 Hz was observed in the 1999-2000 outburst of the same black hole transient from RXTE/PCA data \citep{10.1093/mnras/stac2142}. We hence vary the mass in the above range and find that for the lowest $|m|$  ($m=-2$), leading to best fit, the spin parameter $a$ comes out to be in $0.57 - 0.61$ for the mass $M$ in the range of $5.9 - 9.69M_\odot$. 

The estimated mass range for IGR J17091-3624 is 8.7$ - 15.6 M_\odot$ \citep{Iyer_2015} based on spectro-temporal analysis during the onset of the 2011 outburst. However, \cite{Altamirano_2012} argued that the QPO peak at 164 Hz is a marginal detection, while that at 66 Hz is reliable. In our study, we vary the mass within the above mentioned range, and reproduce both the QPO frequencies. We however find that the entire mass range does not reproduce these QPOs. We have tabulated in Table \ref{tab:BH_param} the result for $m = -3$, which shows that the source could have a fast spinning black hole. The results with $m= -2$, however, argue the spin parameter of $a \sim 0.59  - 0.67$. 
%For this mode we  see however that the entire mass range does not reproduce the observed QPO, 
Thus our results give new constraints on the mass.

The twin HF QPOs of GRS 1915+105 had been reported to be at 67 and 40 Hz \citep{Strohmayer_2001_ApJL554L169}, while 67 Hz alone was reported even earlier \citep{Morgan_Remillard_Greiner1997_ApJ_482_993}. An additional twin peak of $\sim168$ Hz and $\sim113$ Hz were reported further 
\citep{astro-ph/0306213,10.1111/j.1365-2966.2006.10286.x}. 
However, later on, a prominent QPO around 70 Hz has been observed by various groups, see \citealt{10.1093/mnras/stz2143, 10.1093/mnras/stac615} for the observation of AstroSat. Moreover, a very sporadic appearance of peaks at 41 Hz and 34 Hz has also been reported \citep{10.1111/j.1365-2966.2006.10286.x,10.1093/mnras/stt285}. The mass of the source has been estimated to lie in the range of $10.6 - 14.4M_\odot$ \citep{Reid_2014} by independent observation.  However, only the  mass range $11.65 - 14.4M_\odot$ with varying spin parameter $a$ from $0.99$ to $0.8$, for $m = -2$, is successful in reproducing the QPOs at 168 Hz and 113 Hz. For the very similar mass and spin combinations, the pair of 69 Hz and 41.1 Hz
is reproduced by our model but for $m=-6$. However, note that 40 Hz QPOs is observed in the harder state \citep{Strohmayer_2001_ApJL554L169} with the X-ray energy band above 13 KeV, while the current model mostly is in accordance with the Keplerian accretion disk corresponding to the high/soft state (see however Section 7 below). For the mass of $12.4M_\odot$ the spin parameter is estimated to be $a = 0.94$. Thus notice from this analysis that we are further able to constraint the lower bound of the black hole mass, due to a theoretical upper limit of the spin parameter. A more precise measurement of the mass will lead to a more accurate estimate of the spin parameter. 

The mass of XTE J1752-223 has been estimated to be in the range of 3.1$ - 55M_\odot$ \citep{8207506}. However, \cite{Belloni_2012} suggested that the observed HF QPO frequencies, 442.91 Hz and 114.97 Hz, are likely to be  statistical fluctuations, neither in the $3:2$ ratio. Some observations reported its spin to be in the intermediate range with $a \sim 0.52 \pm 0.11$ \citep{10.1111/j.1365-2966.2010.17628.x}. However, a very recent analysis showed that the mass of XTE J1752-223 can be constrained in the range $10 \pm 1.9 M_\odot$ \citep{10.1093/mnras/stab1169}. Nevertheless, had the above mentioned QPO frequencies to be real, we find a better constraint on the mass of $10^{+1.63}_{-1.45} M_\odot$ and the spin parameter in the range of $a\sim 0.6- 0.65$ for $m = -1$ for the best fit of QPOs by our model. 

While reporting four HF QPO frequencies of GX 339-4, 105.35 Hz, 127.7 Hz, 208.44 Hz and 278.64 Hz, \cite{Belloni_2012} argued that all of them are spurious/nonreal. The mass of the black hole GX 339-4 has been varied in $M = 5.8 \pm 0.5 M_\odot$ \citep{Hynes_2003}. If we indeed relied upon the questionable QPO frequencies mentioned above, assuming them to form two pairs as 127.7, 278.64 Hz and 105.35 Hz, 208.44 Hz \citep{Belloni_2012}, we can further constraint the mass of the source. Interestingly, varying the mass within the given range does not provide suitable parameters that could reproduce both frequency pairs assuming a fixed mode ($m$). However considering each pair corresponds to a distinct mode, i.e. mode $m$ changes for different pairs, both the pairs of QPO frequencies are reproduced. We find that the mass in the range of $5.3 - 6.17M_\odot$ with the spin parameter of $a = 0.99 - 0.65$ is able to reproduce first and second frequency pairs for $m = -3\text{ and }-4$ respectively. While \cite{Ludlam_2015} argued for a spin parameter $a>0.97$, our analysis suggests that such a value can only be obtained for $M=6.15 - 6.17M_\odot$, which provides a better constraint on the mass of the black hole GX 339-4.

\section{Explaining other QPOs}

The HF/kHz QPOs need not necessarily always appear in a pair, as evident observationally (see, e.g.. \citealt{Homan_2003, Belloni_2012, Remillard_Morgan99}). We argue that when the peak separation $\Delta\nu=0$, i.e. when both of them coincide with equal amplitude, they exhibit only one QPO. From equation \Eref{nu_s_sigma_nm} for $b_1=-b_2$, we can write 
\begin{equation}
\nu_1-\nu_2=\frac{n-m+1}{b_1}\nu_s.    
\label{nu12mod}
\end{equation}
Further we modify the proposal given by equation (\ref{nuhl}) for general $b_1$ as
\begin{eqnarray}
\nu_h=\nu_1+\frac{m+1}{b_1}\nu_s,\,\,\,
\nu_l=\nu_2+\frac{m}{b_1}\nu_s.
\label{modnuhl}
\end{eqnarray}
Therefore, from equations (\ref{nu12mod}) and (\ref{modnuhl}), we obtain
\begin{equation}
\nu_h-\nu_l=\Delta\nu=\frac{n-m+2}{b_1}\nu_s.    
\end{equation}
Hence, $\Delta\nu=0$, either for higher order nonlinearity  which corresponds to a very large $b_1$ or for $n=m-2$. For the latter, the mode locking has to be led by $\nu_h-\nu_l=0$, rather than $\nu_1-\nu_2=0$.

For the low frequency QPOs, as seen in the hard states of GRS~1915+105 \citep{BelloniGRS}, we must consider nonzero $u_r$ and its radial gradient $\gamma$. This is because a hard accretion state implies significant radial velocity making the flow hot and radiation trapped (see, e.g., \citealt{narayan_yi95,Chakrabarti:1996ns,rajmukh10}). Therefore, any model should include $u_r$. Therefore minimizing $f(\omega)$ in equation \Eref{maxcond} with $\gamma\neq0$, we obtain
\begin{equation}
\omega^2=X-\frac{\gamma^2}{2}\pm\frac{\sqrt{\gamma^4-4X\gamma^2}}{2},
\label{om2}
\end{equation}
where $X=\Omega_\phi(\xi+2\Omega_\phi)+\Omega_r^2$. Now in an advective sub-Keplerian flow $\Omega_\phi\sim r^{-q}$ where $q>1.5$, and $\gamma^2>>\Omega_r^2$. Therefore, from equation (\ref{om2}) we obtain the physically meaningful resonance frequency leading to maximum amplitude of radial and azimuthal perturbations as 
\begin{equation}
\omega=\frac{\sqrt{2}\Omega_r^2}{\gamma},    
\end{equation}
for $q=2$.
For an advective accretion disk with $\alpha$-viscosity=0.01 around a rotating black hole of spin (Kerr) parameter $a=0.99$ \citep{rajmukh10}, the low frequency QPO turns out to be $\gtrsim 0.1$ Hz based on the above model.
Such a QPO frequency is observed in the hard state of GRS~1915+105.

The detailed analyses of above mentioned QPOs based on our model will be presented elsewhere. The present section just imprints that the same model is capable of explaining all kinds of QPOs.

\section{Summary and conclusions}

The origin of QPOs is a longstanding problem, particularly in high energy astrophysics. Over the years, several models have been proposed to enlighten the issue. However, none of them is without any caveat. Moreover, there are several classes of QPOs observed, ranging from mHz to kHz orders, for the black hole, neutron star and white dwarf sources. The question is then, are the origins of all QPOs the same? The HF QPOs, e.g. around black holes, are mostly observed in softer states, while the ones of the order of Hz or 1/10th of Hz are observed in harder states. The present work aims at providing a unified QPO model. We show that HF and kHz QPOs from black holes and neutron stars, respectively, are of the same origin. Other QPOs are also expected to be originated from the same basic mechanism, established in the work.

We have shown explicitly that QPOs of frequency $\sim 100-1000$ Hz are originated from the same mechanism. We have shown that fundamental epicyclic oscillation frequency is modified in the accretion disk compared to what it is for a test particle. When the disk is perturbed by external forces including that due to frame-dragging effects by the spinning compact object at the center, the new modes with frequencies in the combination of fundamental and perturbation frequencies are formed, and they form a resonance around a particular radius. This resonance leads to the locking of new modes which correspond to higher and lower QPO frequencies. 

Now the question is, whether the other QPOs, including the ones having low frequency, are originated from the same mechanism or not.
This is particularly important as some other models, e.g., \citealt{Chakrabarti_2000}, are apparently capable of explaining observed low frequency QPOs in black hole systems. Similarly, the QPOs also appear without their twin. Can the present model be applicable to a single HF/kHz QPO? It is possible to have the latter when $\Delta\nu\sim 0$, i.e. practically one peck is (or both peaks overlap each other), and for higher order nonlinearity or resonance mode-locking at $\nu_h-\nu_l=0$, in place of $\nu_1-\nu_2=0$. The former however corresponds to nonzero radial velocity and its gradient. Indeed the low frequency QPOs appear in the hard state of X-ray binaries when the accretion flow is understood to have a significant advection and be radiation trapped hot. The new resonance condition with the maximum amplitude of the perturbation with a nonzero radial velocity gradient could explain low frequency QPOs. The details will be reported elsewhere. Therefore, in principle, the present model appears to be a truly unified model. 

Our model also predicts and/or constraints the mass of the black holes and neutron stars, particularly if not known for the latter from independent estimate. It also predicts the radius of neutron stars.

It is also possible that the theory of Einstein's general relativity (GR) may not be the ultimate theory of gravity, and there may be some modifications to GR (e.g., \citealt{Upasana15,Kalita18,Kalita_Bani, Das2022}) that could explain the observed varieties of QPOs more judiciously based on the present QPO model. 
%especially asymptotic modification \flag{give all citations}. 
Since QPO is expected to emerge from close to the black hole, such modification to GR can be captured by the QPO frequencies.

\section*{Acknowledgments}
The authors thank Tomaso Belloni for bringing our attention to the latest data, carefully reading the manuscript, comments, and suggestions. Thanks are also due to the referee for comments which have helped to improve clarity including that in the presentation of results in tables. One of the authors (ARD) acknowledges the
financial support from KVPY, DST, India.

\bibliographystyle{aasjournal}
\bibliography{ref.bib}

\begin{thebibliography}{}
\expandafter\ifx\csname natexlab\endcsname\relax\def\natexlab#1{#1}\fi

\bibitem[{Abramowicz {et~al.}(2003{\natexlab{a}})Abramowicz, Bulik, Bursa, \&
  Klu\'{z}niak}]{Abramowicz_2003}
Abramowicz, M.~A., Bulik, T., Bursa, M., \& Klu\'{z}niak, W.
  2003{\natexlab{a}}, A\&A, 404, L21

\bibitem[{Abramowicz {et~al.}(2003{\natexlab{b}})Abramowicz, Karas, Kluzniak,
  Lee, \& Rebusco}]{Abramowicz:2003xy}
Abramowicz, M.~A., Karas, V., Kluzniak, W., Lee, W.~H., \& Rebusco, P.
  2003{\natexlab{b}}, Publ. Astron. Soc. Jap., 55, 466

\bibitem[{Altamirano \& Belloni(2012)}]{Altamirano_2012}
Altamirano, D., \& Belloni, T. 2012, The Astrophysical Journal Letters, 747, L4

\bibitem[{Beer \& Podsiadlowski(2002)}]{Beer_Podsiadlowski_2002}
Beer, M.~E., \& Podsiadlowski, P. 2002, Monthly Notices of the Royal
  Astronomical Society, 331, 351

\bibitem[{Bejger \& Haensel(2002)}]{Bejger_2002}
Bejger, M., \& Haensel, P. 2002, A\&A, 396, 917

\bibitem[{Belloni {et~al.}(2005)Belloni, M\'endez, \&
  Homan}]{belloni2005distribution}
Belloni, T., M\'endez, M., \& Homan, J. 2005, A\&A, 437, 209

\bibitem[{Belloni {et~al.}(2007)Belloni, Méndez, \&
  Homan}]{10.1111/j.1365-2966.2007.11486.x}
Belloni, T., Méndez, M., \& Homan, J. 2007, Monthly Notices of the Royal
  Astronomical Society, 376, 1133

\bibitem[{Belloni {et~al.}(2006{\natexlab{a}})Belloni, Soleri, Casella,
  Méndez, \& Migliari}]{10.1111/j.1365-2966.2006.10286.x}
Belloni, T., Soleri, P., Casella, P., Méndez, M., \& Migliari, S.
  2006{\natexlab{a}}, Monthly Notices of the Royal Astronomical Society, 369,
  305

\bibitem[{Belloni {et~al.}(2006{\natexlab{b}})Belloni, Soleri, Casella,
  Méndez, \& Migliari}]{BelloniGRS}
---. 2006{\natexlab{b}}, Monthly Notices of the Royal Astronomical Society,
  369, 305

\bibitem[{Belloni \& Altamirano(2013)}]{10.1093/mnras/stt285}
Belloni, T.~M., \& Altamirano, D. 2013, Monthly Notices of the Royal
  Astronomical Society, 432, 19

\bibitem[{Belloni {et~al.}(2019)Belloni, Bhattacharya, Caccese, Bhalerao,
  Vadawale, \& Yadav}]{10.1093/mnras/stz2143}
Belloni, T.~M., Bhattacharya, D., Caccese, P., {et~al.} 2019, Monthly Notices
  of the Royal Astronomical Society, 489, 1037

\bibitem[{Belloni {et~al.}(2012)Belloni, Sanna, \& Méndez}]{Belloni_2012}
Belloni, T.~M., Sanna, A., \& Méndez, M. 2012, Monthly Notices of the Royal
  Astronomical Society, 426, 1701

\bibitem[{Blaes {et~al.}(2007)Blaes, Šrámková, Abramowicz, Kluźniak, \&
  Torkelsson}]{Blaes_2007}
Blaes, O.~M., Šrámková, E., Abramowicz, M.~A., Kluźniak, W., \& Torkelsson,
  U. 2007, The Astrophysical Journal, 665, 642

\bibitem[{Blandford \& Znajek(1977)}]{BZ77}
Blandford, R.~D., \& Znajek, R.~L. 1977, \mnras, 179, 433

\bibitem[{Cadez {et~al.}(2008)Cadez, Calvani, \& Kostic}]{Cadez:2008iv}
Cadez, A., Calvani, M., \& Kostic, U. 2008, Astron. Astrophys., 487, 527

\bibitem[{Chakrabarti(1996)}]{Chakrabarti:1996ns}
Chakrabarti, S.~K. 1996, Astrophys. J., 464, 664

\bibitem[{Chakrabarti \& Manickam(2000)}]{Chakrabarti_2000}
Chakrabarti, S.~K., \& Manickam, S.~G. 2000, The Astrophysical Journal, 531,
  L41

\bibitem[{Connors {et~al.}(1980)Connors, Piran, \& Stark}]{1980ApJ...235..224C}
Connors, P.~A., Piran, T., \& Stark, R.~F. 1980, \apj, 235, 224

\bibitem[{Cook {et~al.}(1994)Cook, Shapiro, \& Teukolsky}]{1994ApJ}
Cook, G.~B., Shapiro, S.~L., \& Teukolsky, S.~A. 1994, Astrophysical Journal,
  424, 823

\bibitem[{Das \& Mukhopadhyay(2022)}]{Das2022}
Das, A.~R., \& Mukhopadhyay, B. 2022, The European Physical Journal C, 82, 939

\bibitem[{Das \& Mukhopadhyay(2015)}]{Upasana15}
Das, U., \& Mukhopadhyay, B. 2015, Journal of Cosmology and Astroparticle
  Physics, 2015, 045

\bibitem[{Davis {et~al.}(2005)Davis, Blaes, Hubeny, \& Turner}]{Davis_2005}
Davis, S.~W., Blaes, O.~M., Hubeny, I., \& Turner, N.~J. 2005, The
  Astrophysical Journal, 621, 372

\bibitem[{Debnath {et~al.}(2021)Debnath, Chatterjee, Chatterjee, Jana, \&
  Chakrabarti}]{10.1093/mnras/stab1169}
Debnath, D., Chatterjee, K., Chatterjee, D., Jana, A., \& Chakrabarti, S.~K.
  2021, Monthly Notices of the Royal Astronomical Society, 504, 4242

\bibitem[{Dovčiak {et~al.}(2004)Dovčiak, Karas, \& Yaqoob}]{Dovčiak_2004}
Dovčiak, M., Karas, V., \& Yaqoob, T. 2004, The Astrophysical Journal
  Supplement Series, 153, 205

\bibitem[{Franchini {et~al.}(2016)Franchini, Motta, \&
  Lodato}]{10.1093/mnras/stw3363}
Franchini, A., Motta, S.~E., \& Lodato, G. 2016, Monthly Notices of the Royal
  Astronomical Society, 467, 145

\bibitem[{Germana {et~al.}(2009)Germana, Kostic, Cadez, \&
  Calvani}]{Germana:2009ce}
Germana, C., Kostic, U., Cadez, A., \& Calvani, M. 2009, AIP Conf. Proc., 1126,
  367

\bibitem[{Greiner {et~al.}(2001)Greiner, Cuby, \& McCaughrean}]{Greiner2001}
Greiner, J., Cuby, J.~G., \& McCaughrean, M.~J. 2001, Nature, 414, 522,
  35107019

\bibitem[{Güver {et~al.}(2010)Güver, Özel, Cabrera-Lavers, \&
  Wroblewski}]{G_ver_2010}
Güver, T., Özel, F., Cabrera-Lavers, A., \& Wroblewski, P. 2010, The
  Astrophysical Journal, 712, 964

\bibitem[{Homan {et~al.}(2003)Homan, Klein-Wolt, Rossi, Miller, Wijnands,
  Belloni, van~der Klis, \& Lewin}]{Homan_2003}
Homan, J., Klein-Wolt, M., Rossi, S., {et~al.} 2003, The Astrophysical Journal,
  586, 1262

\bibitem[{Homan {et~al.}(2005)Homan, Miller, Wijnands, van~der Klis, Belloni,
  Steeghs, \& Lewin}]{2005ApJ...623..383H}
Homan, J., Miller, J.~M., Wijnands, R., {et~al.} 2005, \apj, 623, 383

\bibitem[{Hynes {et~al.}(2003)Hynes, Steeghs, Casares, Charles, \&
  O'Brien}]{Hynes_2003}
Hynes, R.~I., Steeghs, D., Casares, J., Charles, P.~A., \& O'Brien, K. 2003,
  The Astrophysical Journal, 583, L95

\bibitem[{Iyer {et~al.}(2015)Iyer, Nandi, \& Mandal}]{Iyer_2015}
Iyer, N., Nandi, A., \& Mandal, S. 2015, The Astrophysical Journal, 807, 108

\bibitem[{Jonker {et~al.}(2002)Jonker, M{\'{e} }ndez, \& van~der
  Klis}]{Jonker_2002}
Jonker, P.~G., M{\'{e} }ndez, M., \& van~der Klis, M. 2002, Monthly Notices of
  the Royal Astronomical Society, 336, L1

\bibitem[{Kalita \& Mukhopadhyay(2018)}]{Kalita18}
Kalita, S., \& Mukhopadhyay, B. 2018, Journal of Cosmology and Astroparticle
  Physics, 2018, 007–007

\bibitem[{Kalita \& Mukhopadhyay(2019)}]{Kalita_Bani}
---. 2019, The European Physical Journal C, 79,
  doi:10.1140/epjc/s10052-019-7396-x

\bibitem[{Kluzniak \& Abramowicz(2002)}]{Kluzniak:2002bb}
Kluzniak, W., \& Abramowicz, M.~A. 2002, arXiv:astro-ph/0203314

\bibitem[{Kluźniak {et~al.}(2004)Kluźniak, Abramowicz, Kato, Lee, \&
  Stergioulas}]{Kluźniak_2004}
Kluźniak, W., Abramowicz, M.~A., Kato, S., Lee, W.~H., \& Stergioulas, N.
  2004, The Astrophysical Journal, 603, L89

\bibitem[{Kostic {et~al.}(2009)Kostic, Cadez, Calvani, \&
  Gomboc}]{Kostic:2009hp}
Kostic, U., Cadez, A., Calvani, M., \& Gomboc, A. 2009, Astron. Astrophys.,
  496, 307

\bibitem[{Kotrlov\'a {et~al.}(2020)Kotrlov\'a, \v{S}r\'amkov\'a, T\"or\"ok,
  Goluchov\'a, Hor\'ak, Straub, Lancov\'a, Stuchl\'\i{}k, \&
  Abramowicz}]{Kotrlova:2020pqy}
Kotrlov\'a, A., \v{S}r\'amkov\'a, E., T\"or\"ok, G., {et~al.} 2020, Astron.
  Astrophys., 643, A31

\bibitem[{Kulkarni {et~al.}(2011)Kulkarni, Penna, Shcherbakov, Steiner,
  Narayan, Sądowski, Zhu, McClintock, Davis, \&
  McKinney}]{10.1111/j.1365-2966.2011.18446.x}
Kulkarni, A.~K., Penna, R.~F., Shcherbakov, R.~V., {et~al.} 2011, Monthly
  Notices of the Royal Astronomical Society, 414, 1183

\bibitem[{Lamb \& Miller(2001)}]{Lamb_2001}
Lamb, F.~K., \& Miller, M.~C. 2001, The Astrophysical Journal, 554, 1210

\bibitem[{Landau \& Lifshitz(1976)}]{LL1976}
Landau, L.~D., \& Lifshitz, E.~M. 1976, Mechanics (Oxford: Pergamon)

\bibitem[{Li {et~al.}(2005)Li, Zimmerman, Narayan, \& McClintock}]{Li_2005}
Li, L.-X., Zimmerman, E.~R., Narayan, R., \& McClintock, J.~E. 2005, The
  Astrophysical Journal Supplement Series, 157, 335

\bibitem[{Lightman \& Shapiro(1975)}]{1975ApJ...198L..73L}
Lightman, A.~P., \& Shapiro, S.~L. 1975, \apjl, 198, L73

\bibitem[{Ludlam {et~al.}(2015)Ludlam, Miller, \& Cackett}]{Ludlam_2015}
Ludlam, R.~M., Miller, J.~M., \& Cackett, E.~M. 2015, The Astrophysical
  Journal, 806, 262

\bibitem[{Majumder {et~al.}(2022)Majumder, Sreehari, Aftab, Katoch, Das, \&
  Nandi}]{10.1093/mnras/stac615}
Majumder, S., Sreehari, H., Aftab, N., {et~al.} 2022, Monthly Notices of the
  Royal Astronomical Society, 512, 2508

\bibitem[{Markwardt {et~al.}(1999)Markwardt, Strohmayer, \&
  Swank}]{Markwardt_1999}
Markwardt, C.~B., Strohmayer, T.~E., \& Swank, J.~H. 1999, The Astrophysical
  Journal, 512, L125

\bibitem[{Mauche(2002)}]{Mauche_2002}
Mauche, C.~W. 2002, The Astrophysical Journal, 580, 423

\bibitem[{McClintock \& Remillard(2006)}]{astro-ph/0306213}
McClintock, J.~E., \& Remillard, R.~A. 2006, Black Hole Binaries, ed. W.~Lewin
  \& M.~van~der Klis, Cambridge Astrophysics (Cambridge University Press),
  157--206

\bibitem[{Messenger {et~al.}(2015)Messenger, Bulten, Crowder, Dergachev,
  Galloway, Goetz, Jonker, Lasky, Meadors, Melatos, Premachandra, Riles,
  Sammut, Thrane, Whelan, \& Zhang}]{PhysRevD.92.023006}
Messenger, C., Bulten, H.~J., Crowder, S.~G., {et~al.} 2015, Phys. Rev. D, 92,
  023006

\bibitem[{Miller {et~al.}(2006)Miller, Raymond, Homan, Fabian, Steeghs,
  Wijnands, Rupen, Charles, van~der Klis, \& Lewin}]{Miller_2006}
Miller, J.~M., Raymond, J., Homan, J., {et~al.} 2006, The Astrophysical
  Journal, 646, 394

\bibitem[{Molla {et~al.}(2017)Molla, Chakrabarti, Debnath, \&
  Mondal}]{Molla:2016mip}
Molla, A.~A., Chakrabarti, S.~K., Debnath, D., \& Mondal, S. 2017, Astrophys.
  J., 834, 88

\bibitem[{Morgan {et~al.}(1997)Morgan, Remillard, \&
  Greiner}]{Morgan_Remillard_Greiner1997_ApJ_482_993}
Morgan, E.~H., Remillard, R.~A., \& Greiner, J. 1997, The Astrophysical
  Journal, 482, 993

\bibitem[{Motch {et~al.}(1983)Motch, Ricketts, Page, Ilovaisky, \&
  Chevalier}]{motch1983simultaneous}
Motch, C., Ricketts, M., Page, C., Ilovaisky, S., \& Chevalier, C. 1983,
  Astronomy and Astrophysics, vol. 119, no. 2, Mar. 1983, p. 171-176. Research
  supported by the Science and Engineering Research Council of England., 119,
  171

\bibitem[{Motta {et~al.}(2012)Motta, Homan, Muñoz-Darias, Casella, Belloni,
  Hiemstra, \& Méndez}]{10.1111/j.1365-2966.2012.22037.x}
Motta, S., Homan, J., Muñoz-Darias, T., {et~al.} 2012, Monthly Notices of the
  Royal Astronomical Society, 427, 595

\bibitem[{Motta {et~al.}(2022)Motta, Belloni, Stella, Pappas, Casares,
  Muñoz-Darias, Torres, \& Yanes-Rizo}]{10.1093/mnras/stac2142}
Motta, S.~E., Belloni, T., Stella, L., {et~al.} 2022, Monthly Notices of the
  Royal Astronomical Society, 517, 1469

\bibitem[{Motta {et~al.}(2013)Motta, Belloni, Stella, Muñoz-Darias, \&
  Fender}]{10.1093/mnras/stt2068}
Motta, S.~E., Belloni, T.~M., Stella, L., Muñoz-Darias, T., \& Fender, R.
  2013, Monthly Notices of the Royal Astronomical Society, 437, 2554

\bibitem[{Mukhopadhyay(2009)}]{Mukhopadhyay_2009}
Mukhopadhyay, B. 2009, The Astrophysical Journal, 694, 387

\bibitem[{Mukhopadhyay {et~al.}(2003)Mukhopadhyay, Ray, Dey, \&
  Dey}]{BMApJL2003}
Mukhopadhyay, B., Ray, S., Dey, J., \& Dey, M. 2003, The Astrophysical Journal,
  584, L83

\bibitem[{Méndez \& van~der Klis(1999)}]{Méndez_1999}
Méndez, M., \& van~der Klis, M. 1999, The Astrophysical Journal, 517, L51

\bibitem[{Méndez \& van~der Klis(2000)}]{10.1046/j.1365-8711.2000.03788.x}
---. 2000, Monthly Notices of the Royal Astronomical Society, 318, 938

\bibitem[{Méndez {et~al.}(1998)Méndez, van~der Klis, Wijnands, Ford, van
  Paradijs, \& Vaughan}]{Méndez_1998}
Méndez, M., van~der Klis, M., Wijnands, R., {et~al.} 1998, The Astrophysical
  Journal, 505, L23

\bibitem[{Nakahira {et~al.}(2012)Nakahira, Koyama, Ueda, Yamaoka, Sugizaki,
  Mihara, Matsuoka, Yoshida, Makishima, Ebisawa, Kubota, Yamada, Negoro, Hiroi,
  Ishikawa, Kawai, Kimura, Kitayama, Kohama, Matsumura, Morii, Nakajima,
  Serino, Shidatsu, Sootome, Sugimori, Suwa, Tomida, Tsuboi, Tsunemi, Ueno,
  Usui, Yamamoto, Yamazaki, Tashiro, Terada, \& Seta}]{8207506}
Nakahira, S., Koyama, S., Ueda, Y., {et~al.} 2012, Publications of the
  Astronomical Society of Japan, 64, 13

\bibitem[{Narayan \& Yi(1995)}]{narayan_yi95}
Narayan, R., \& Yi, I. 1995, \apj, 452, 710

\bibitem[{Nayfeh \& Mook(1979)}]{nayfeh1979nonlinear}
Nayfeh, A., \& Mook, D. 1979, Nonlinear Oscillations, Pure and Applied
  Mathematics: A Wiley Series of Texts, Monographs and Tracts (Wiley)

\bibitem[{Nowak {et~al.}(1997)Nowak, Wagoner, Begelman, \& Lehr}]{Nowak:1996hg}
Nowak, M.~A., Wagoner, R.~V., Begelman, M.~C., \& Lehr, D.~E. 1997, Astrophys.
  J. Lett., 477, L91

\bibitem[{Orosz \& Bailyn(1997)}]{Orosz_1997}
Orosz, J.~A., \& Bailyn, C.~D. 1997, The Astrophysical Journal, 477, 876

\bibitem[{Orosz {et~al.}(2002)Orosz, Groot, van~der Klis, McClintock, Garcia,
  Zhao, Jain, Bailyn, \& Remillard}]{Orosz_2002}
Orosz, J.~A., Groot, P.~J., van~der Klis, M., {et~al.} 2002, The Astrophysical
  Journal, 568, 845

\bibitem[{P\'etri(2005)}]{refId0}
P\'etri, J. 2005, A\&A, 439, 443

\bibitem[{Rajesh \& Mukhopadhyay(2010)}]{rajmukh10}
Rajesh, S.~R., \& Mukhopadhyay, B. 2010, Monthly Notices of the Royal
  Astronomical Society, 402, 961

\bibitem[{Rebusco(2004)}]{Rebusco:2004ba}
Rebusco, P. 2004, Publ. Astron. Soc. Jap., 56, 553

\bibitem[{Rebusco(2008)}]{REBUSCO2008855}
---. 2008, New Astronomy Reviews, 51, 855, jean-Pierre Lasota, X-ray Binaries,
  Accretion Disks and Compact Stars

\bibitem[{Reid {et~al.}(2014)Reid, McClintock, Steiner, Steeghs, Remillard,
  Dhawan, \& Narayan}]{Reid_2014}
Reid, M.~J., McClintock, J.~E., Steiner, J.~F., {et~al.} 2014, The
  Astrophysical Journal, 796, 2

\bibitem[{Reis {et~al.}(2011)Reis, Miller, Fabian, Cackett, Maitra, Reynolds,
  Rupen, Steeghs, \& Wijnands}]{10.1111/j.1365-2966.2010.17628.x}
Reis, R.~C., Miller, J.~M., Fabian, A.~C., {et~al.} 2011, Monthly Notices of
  the Royal Astronomical Society, 410, 2497

\bibitem[{Remillard \&
  McClintock(2006)}]{doi:10.1146/annurev.astro.44.051905.092532}
Remillard, R.~A., \& McClintock, J.~E. 2006, Annual Review of Astronomy and
  Astrophysics, 44, 49

\bibitem[{Remillard \& Morgan(1999)}]{Remillard_Morgan99}
Remillard, R.~A., \& Morgan, E.~H. 1999, in American Astronomical Society
  Meeting Abstracts, Vol. 195, American Astronomical Society Meeting Abstracts,
  37.02

\bibitem[{Remillard {et~al.}(2002)Remillard, Muno, McClintock, \&
  Orosz}]{Remillard_2002}
Remillard, R.~A., Muno, M.~P., McClintock, J.~E., \& Orosz, J.~A. 2002, The
  Astrophysical Journal, 580, 1030

\bibitem[{Reynolds \& Fabian(2008)}]{Reynolds_2008_1}
Reynolds, C.~S., \& Fabian, A.~C. 2008, The Astrophysical Journal, 675, 1048

\bibitem[{Reynolds \& Nowak(2003)}]{REYNOLDS2003389}
Reynolds, C.~S., \& Nowak, M.~A. 2003, Physics Reports, 377, 389

\bibitem[{Rezzolla {et~al.}(2003)Rezzolla, Yoshida, Maccarone, \&
  Zanotti}]{10.1046/j.1365-8711.2003.07018.x}
Rezzolla, L., Yoshida, S., Maccarone, T.~J., \& Zanotti, O. 2003, Monthly
  Notices of the Royal Astronomical Society, 344, L37

\bibitem[{Shahbaz {et~al.}(1999)Shahbaz, van~der Hooft, Casares, Charles, \&
  van Paradijs}]{10.1046/j.1365-8711.1999.02481.x}
Shahbaz, T., van~der Hooft, F., Casares, J., Charles, P.~A., \& van Paradijs,
  J. 1999, Monthly Notices of the Royal Astronomical Society, 306, 89

\bibitem[{Smith {et~al.}(1997)Smith, Morgan, \& Bradt}]{Smith_1997}
Smith, D.~A., Morgan, E.~H., \& Bradt, H. 1997, The Astrophysical Journal, 479,
  L137

\bibitem[{Steeghs \& Casares(2002)}]{Steeghs_2002_1}
Steeghs, D., \& Casares, J. 2002, The Astrophysical Journal, 568, 273

\bibitem[{Steiner {et~al.}(2011{\natexlab{a}})Steiner, McClintock, \&
  Reid}]{Steiner_2011_2}
Steiner, J.~F., McClintock, J.~E., \& Reid, M.~J. 2011{\natexlab{a}}, The
  Astrophysical Journal, 745, L7

\bibitem[{Steiner {et~al.}(2011{\natexlab{b}})Steiner, Reis, McClintock,
  Narayan, Remillard, Orosz, Gou, Fabian, \& Torres}]{Steiner_2011}
Steiner, J.~F., Reis, R.~C., McClintock, J.~E., {et~al.} 2011{\natexlab{b}},
  Monthly Notices of the Royal Astronomical Society, 416, 941

\bibitem[{Stella \& Vietri(1998)}]{Stella:1997tc}
Stella, L., \& Vietri, M. 1998, Astrophys. J. Lett., 492, L59

\bibitem[{Stella {et~al.}(1999)Stella, Vietri, \& Morsink}]{Stella:1999sj}
Stella, L., Vietri, M., \& Morsink, S. 1999, Astrophys. J. Lett., 524, L63

\bibitem[{Strohmayer(2001)}]{Strohmayer_2001_ApJL554L169}
Strohmayer, T.~E. 2001, The Astrophysical Journal, 552, L49

\bibitem[{Titarchuk \& Wood(2002)}]{Titarchuk_2002}
Titarchuk, L., \& Wood, K. 2002, The Astrophysical Journal, 577, L23

\bibitem[{Torok {et~al.}(2010)Torok, Bakala, Sramkova, Stuchlik, \&
  Urbanec}]{Torok:2010rk}
Torok, G., Bakala, P., Sramkova, E., Stuchlik, Z., \& Urbanec, M. 2010,
  Astrophys. J., 714, 748

\bibitem[{Torok {et~al.}(2011)Torok, Kotrlova, Sramkova, \&
  Stuchlik}]{Torok:2011qy}
Torok, G., Kotrlova, A., Sramkova, E., \& Stuchlik, Z. 2011, Astron.
  Astrophys., 531, A59

\bibitem[{Török {et~al.}(2015)Török, Goluchová, Horák, Šrámková,
  Urbanec, Pecháček, \& Bakala}]{10.1093/mnrasl/slv196}
Török, G., Goluchová, K., Horák, J., {et~al.} 2015, Monthly Notices of the
  Royal Astronomical Society: Letters, 457, L19

\bibitem[{van~der Klis(2000)}]{van_der_Klis_2000}
van~der Klis, M. 2000, Annual Review of Astronomy and Astrophysics, 38, 717

\bibitem[{van~der Klis {et~al.}(1985)van~der Klis, Jansen, van Paradijs, Lewin,
  van~den Heuvel, Trümper, \& Sztajno}]{vanderKlis1985}
van~der Klis, M., Jansen, F., van Paradijs, J., {et~al.} 1985, Nature, 316, 225

\bibitem[{van Straaten {et~al.}(2002)van Straaten, van~der Klis, di~Salvo, \&
  Belloni}]{van_Straaten_2002}
van Straaten, S., van~der Klis, M., di~Salvo, T., \& Belloni, T. 2002, The
  Astrophysical Journal, 568, 912

\bibitem[{Woudt \& Warner(2002)}]{2002MNRAS.333..411W}
Woudt, P.~A., \& Warner, B. 2002, \mnras, 333, 411

\bibitem[{Yanes-Rizo {et~al.}(2022)Yanes-Rizo, Torres, Casares, Motta,
  Muñoz-Darias, Rodríguez-Gil, Armas Padilla, Jiménez-Ibarra, Jonker,
  Corral-Santana, \& Fender}]{10.1093/mnras/stac2719}
Yanes-Rizo, I.~V., Torres, M. A.~P., Casares, J., {et~al.} 2022, Monthly
  Notices of the Royal Astronomical Society, 517, 1476

\bibitem[{Zhang {et~al.}(1997)Zhang, Cui, \& Chen}]{Zhang_1997}
Zhang, S.~N., Cui, W., \& Chen, W. 1997, The Astrophysical Journal, 482, L155

\bibitem[{Özel {et~al.}(2012)Özel, Gould, \& Güver}]{Ozel_2012}
Özel, F., Gould, A., \& Güver, T. 2012, The Astrophysical Journal, 748, 5

\end{thebibliography}

\end{document}